\documentclass[journal]{IEEEtran}

\usepackage[cmex10]{amsmath}
\usepackage{graphicx}
\usepackage{cite}
\usepackage{epstopdf}
\usepackage{algorithm}
\usepackage{algorithmic}
\usepackage[caption = false]{subfig}
\usepackage{booktabs}
\usepackage{color}

\begin{document}
\title{Energy Efficiency Maximization of Full-Duplex Two-Way Relay with Non-ideal Power Amplifiers and Non-negligible Circuit Power}

\author{\IEEEauthorblockN{Qimei~Cui,~\IEEEmembership{Senior~Member,~IEEE}, Yuhao~Zhang, Wei~Ni,~\IEEEmembership{Senior~Member,~IEEE}, Mikko~Valkama,~\IEEEmembership{Senior~Member,~IEEE}, and Riku~J\"{a}ntti,~\IEEEmembership{Senior~Member,~IEEE}}

\thanks{This work was supported in part by the National Nature Science Foundation of China Project under Grant 61471058, in part by the Key National Science Foundation of China under Grant 61461136002 and 61231009, in part by the Hong Kong, Macao and Taiwan Science and Technology Cooperation Projects under Grant 2016YFE0122900, and in part by the 111 Project of China under Grant B16006.

Q. Cui and Y. Zhang are with School of Information and Communication Engineering, Beijing University of Posts and Telecommunications, 100876, Beijing, China (e-mail: cuiqimei@bupt.edu.cn).

W. Ni is with Commonwealth Scientific and Industrial Research Organisation (CSIRO), Canberra, ACT 2600, Australia (e-mail: wei.ni@csiro.au).

M. Valkama is with the Department of Electronics and Communications Engineering, Tampere University of Technology, 33101, Tampere, Finland (e-mail: mikko.e.valkama@tut.fi).

R. J\"{a}ntti is with the Department of Communications and Networking, Aalto University, 02150, Espoo, Finland (e-mail: riku.jantti@aalto.fi).}}

\maketitle

\begin{abstract}
In this paper, we maximize the energy efficiency~(EE) of full-duplex~(FD) two-way relay~(TWR) systems under non-ideal power amplifiers (PAs) and non-negligible transmission-dependent circuit power. We start with the case where only the relay operates full duplex and two timeslots are required for TWR. Then, we extend to the advanced case, where the relay and the two nodes all operate full duplex, and accomplish TWR in a single timeslot. In both cases, we establish the intrinsic connections between the optimal transmit powers and durations, based on which the original non-convex EE maximization can be convexified and optimally solved. Simulations show the superiority of FD-TWR in terms of EE, especially when traffic demand is high. The simulations also reveal that the maximum EE of FD-TWR is more sensitive to the PA efficiency, than it is to self-cancellation. The full FD design of FD-TWR is susceptible to traffic imbalance, while the design with only the relay operating in the FD mode exhibits strong tolerance.
\end{abstract}

\begin{IEEEkeywords}
Energy efficiency, full-duplex, non-ideal power amplifier, non-negligible circuit power, two-way relay.
\end{IEEEkeywords}

\IEEEpeerreviewmaketitle

\section{Introduction}
Full duplex~(FD) has recently been extensively studied due to significant advance on self-cancellation techniques~\cite{Experiment-Driven,Reference34,Reference13,Reference32,Reference10,Reference2}. The self-interference from the transmit antenna(s) of a node to the receive antenna(s) of the same node can be suppressed by means of passive suppression (PS) such as antenna separation~\cite{Reference13,Reference32,Reference2}, and active cancellation~(AC) such as radio-frequency circuit designs and digital signal processing~\cite{Experiment-Driven,Reference13,Reference10,Reference2}. By collectively using these techniques, a self-cancellation of 80 dB, comprising 40 dB from PS and 40 dB from AC, has been recorded~\cite{Reference13,Reference2}.

Two-way relay~(TWR) has also drawn extensive recent interest, where advanced digital signal processing has been developed to minimize the number of timeslots required for two mutually hidden nodes to exchange data via a shared half-duplex~(HD) relay~\cite{Reference49,PLNC&NC}. Physical-layer network coding~(PNC) has been proposed to achieve the minimum number of two slots under HD settings~\cite{PLNC&NC}. TWR also has the advantage of improving energy efficiency~(EE), as extensively studied in~\cite{Reference5,Reference8,Reference45,Reference48}.

There is an opportunity to join the advantages of FD and TWR, referred to as FD-TWR, to improve the spectral efficiency~(SE) and EE of TWR. Practical conditions need to be taken into account, such as limited power supply in mobile nodes, non-ideal power amplifier~(PA) and non-negligible circuit power consumption. To this end, the maximization of EE is of practical interest, especially when the relay and hidden nodes are mobile nodes and do not have persistent power supply. Despite the EE maximization has been extensively studied in HD relay-assisted wireless networks~\cite{Reference5,Reference8,Reference45,Reference48}, few works have been carried out on HD-TWR under non-ideal PA and non-negligible circuit power, not to mention FD-TWR. To the best of our knowledge, the collective impact of non-ideal PAs, non-negligible circuit power, and self-interference on FD-TWR has not been studied heretofore.

In most cases, the non-ideal PAs and non-negligible circuit power are non-linear to the transmit power. Under prevailing PA models, such as traditional PA~(TPA)~\cite{EEdelay}, the PA power is non-linear to the transmit power. In contrast, the non-negligible circuit power is non-linear and non-convex, since it is typically linear to the data rate which takes the logarithm of the transmit power. The non-ideal PAs and non-negligible circuit power necessitate new optimization variables, i.e., transmit durations, apart from the transmit powers. Particularly, if the nodes transmit excessively long, the circuit energy consumption on reception would increase. If the nodes transmit too short, the transmit powers would become excessively high, while the standby energy of the circuits would grow. The EE would degrade in either of these cases.

In this paper, the EE of FD-TWR is maximized under non-ideal PAs and non-negligible rate-dependent circuit power. We start with the case where only the relay is FD enabled and can transmit to one of two end nodes while receiving from the other. Two timeslots are required for TWR (the case is referred to as FD-TWR-2TS). The EE maximization is formulated with intractable non-convex constraints resulting from self-interference at the FD relay. The necessary conditions of the optimal solution are derived to convert the original non-convex problem to a convex problem with respect to~(w.r.t) the transmit durations. We further extend the necessary conditions to the advanced FD-TWR design, referred to as FD-TWR-1TS, where the relay and the two end nodes all operate in the FD mode, and exploit PNC to accomplish TWR within a single timeslot. The EE maximization of FD-TWR-1TS is also proved to be non-convex. The necessary conditions facilitate convexifying this problem while preserving optimality. Corroborated by extensive simulations, the convex reformulation of both FD-TWR-2TS and FD-TWR-1TS can be readily implemented using standard CVX tools.

The contributions of this paper are summarized as follows.
\begin{itemize}
  \item The necessary conditions of the most energy efficient schedule for FD-TWR, which are key to convexify the problem and preserve optimality;
  \item Convexification of the EE maximization for FD-TWR under non-ideal PAs and non-negligible circuit powers, which enables efficient solvers to solve the problem with guaranteed convergence and optimality;
  \item Important findings: $(a)$ the EE of FD-TWR is more sensitive to the PA efficiency, than it is to self-cancellation; $(b)$ FD-TWR-1TS is susceptible to traffic imbalance, while FD-TWR-2TS exhibits strong tolerance; $(c)$ FD-TWR is superior to HD-TWR in terms of EE in the case of high data rate demand.
\end{itemize}

The rest of the paper is organized as follows. Section~\ref{Sec2} surveys the related works. Section~\ref{Sec3} presents the system model. The EE maximization problems of FD-TWR-2TS and FD-TWR-1TS are formulated and solved in Sections~\ref{Sec4} and~\ref{Sec5}, respectively. Simulation results are provided in Section~\ref{Sec6}, followed by a conclusion in Section~\ref{Sec7}.
\section{Related Work}\label{Sec2}
Earlier works~\cite{Reference5,Reference8,Reference45,Reference48} maximized the EE of HD-TWR under the assumption of ideal PA model and negligible circuit power. PAs are typically non-ideal with the PA efficiency changing with the transmit power, and the circuit powers (i.e., baseband processing, radio-frequency (RF) generation and the circuit standby power etc.) are non-negligible, ranging from tens to hundreds of milliWatts~\cite{Reference43,Reference46}. In~\cite{Reference41}, under the assumption of ideal PA, non-negligible circuit power consumption was considered when the EE of amplify-and-forward~(AF) HD-TWR was maximized. The EE was quasi-concave over the transmit power, and maximized by using the Dinkelbachs method. In~\cite{Reference47}, the aggregated utility of EE and proportional fairness were maximized for OFDMA-based HD-TWR where the circuit power in active sub-channels was evaluated for energy consumption. In~\cite{Reference24}, the EE maximization of two timeslots decode-and-forward~(DF) HD-TWR was formulated under the non-ideal TPA model and solved using a standard CVX toolbox. In~\cite{Reference43}, the EE maximization was generalized under a variety of non-ideal PA models and HD-TWR strategies ({e.g.}, two and three timeslots TWR). The generalized problems were converted to problems with convex structures in the vicinities of the optimum, and solved with guaranteed optimality.

Only a handful of works have been on focused the EE maximization of FD-TWR, typically under the assumptions of ideal PAs and/or constant/transmission-independent circuit power. In~\cite{Reference12}, a non-convex problem was cast to maximize the EE of AF FD-TWR given a SE requirement. A suboptimal solution was developed, comprising two alternating steps. In~\cite{Reference42}, four typical power-scaling schemes were proposed to improve the EE of multi-pair AF FD-TWR with a massive antenna array at the relay. With the number of antennas going to infinity, the asymptotic SE and EE were derived. In~\cite{latticecode}, a new Lattice code with structured binning was designed for FD-TWR, where the relay only quantizes RF combined signals of the end nodes by a constellation lattice and forwards the quantized signals. The end nodes can recover their desired signals, exploiting the meticulously designed binning.

Most existing works on the EE of FD have been on one-way relaying, typically under the assumption of ideal PAs and/or constant circuit power consumption. In~\cite{Reference35}, the closed-form expressions for the achievable ergodic rates were derived for dual-hop FD massive MIMO AF systems. In~\cite{Reference14}, discrete stochastic optimization and the Dinkelbach method were taken in an alternating manner to maximize the EE of a multi-cell OFDMA network with shared FD relays in coverage overlapping areas. In~\cite{Reference36}, a selective DF protocol was proposed to select FD nodes for cooperative relay, given a required outage threshold. In~\cite{Reference13}, a comparison study was conducted on the EE of FD and HD dual-hop AF systems. Opportunistic relaying mode selection in coupling with transmission power adaptation was developed to maximize the EE. In~\cite{HD-FD-buffer}, HD relay was shown to outperform FD relay in terms of throughput under a DF mode at a cost of the increased relay buffer. However, this conclusion is inapplicable to advanced PNC, such as the Lattice code (as considered in this paper), where signals are only decoded at the destination. The relay just quantizes and forwards the signals. The requirement of the relay buffer is constant, depending on the quantization delay.

In different yet relevant contexts, the impact of non-ideal PAs or non-negligible circuit power was evaluated in other communication systems. In~\cite{EEdelay}, the EE-delay tradeoff of a proportional-fair downlink cellular system was studied under non-ideal PAs. In~\cite{EEmetric}, the power allocation was optimized to maximize a bits-per-Joule EE in the conventional single-hop frequency-selective channels under the assumption of non-negligible constant circuit power. In~\cite{Reference21}, a ¡°string tautening¡± algorithm was proposed to produce the most energy-efficient schedule for delay-limited traffic, first under the assumption of negligible circuit power, and then extended to non-negligible constant circuit powers~\cite{circuitp1} and energy-harvesting communications~\cite{circuitp2}. In~\cite{Reference11}, the rate region of a FD OFDM link was maximized under non-ideal transceivers by modelling the self-interference as an additive error vector magnitude~(EVM) noise which was decomposed into the equivalent noises at the two ends of the FD link. The equivalent SINRs of the two link directions were separately formulated and jointly maximized by taking the sub-gradient method. However, the results of these works are not applicable to FD-TWR, due to distinctive system architectures.
\section{System Model}\label{Sec3}
Consider an in-band FD-TWR point-to-point wireless network~\cite{P2P_1,P2P_2}, where there are two end nodes, termed $A$ and $B$, and a relay node, termed $R$, indicated by the subscripts ``$_a$'', ``$_b$'' and ``$_r$'', respectively. Node $R$ lies between nodes $A$ and $B$. All the nodes are equipped with directional antennas. Nodes $A$ and $B$ do not have a direct link, and need to exchange traffic through node $R$ at the same frequency~\cite{Reference12,Reference42}. This scenario is typical in practice, since the three nodes are unaligned in many cases and the directivity of the directional antennas prevents nodes $A$ and $B$ from having a direct link. In many other cases, nodes $A$ and $B$ are too far away and beyond reach. The three nodes can operate in different modes, depending on the FD-TWR strategies specified shortly. The system bandwidth is $W$. Flat-fading channels are assumed with complex channel coefficients $h_{i,j} \, (i,j \in \{a, b, r\})$ from node $i$ to $j$. Particularly, $h_{i,i}$ is the complex channel coefficient of the self-interference channel at node $i$, from the transmit antenna of the node to its own receive antenna. The additive white Gaussian noise~(AWGN) has zero mean and variance $\sigma_j^2$, denoted by $n_j \sim \mathcal{CN} (0,\sigma_j^2)$.
\subsection{Full-Duplex and Self-interference Cancellation}\label{Sec3:1}
Under the FD mode, the self-interference can be partly suppressed by PS and/or AC. Generally, PS, on its own, can account for over 40 dB reduction of self-interference. Working together with AC, it can suppress up to 80 dB. The joint use of PS and AC is referred to as PSAC~\cite{Reference32}. We denote the ratio of the self-interference before and after self-cancellation as $\alpha_{PS}$ and $\alpha_{PSAC}$ under PS and PSAC, respectively.

The residual self-interference, after self-cancellation, is typically modeled as a zero-mean Gaussian random variable~\cite{ResidualNoise,ResidualNoise1}. The residual self-interference at node $i$ is $e_i = \sqrt {P_i} {\tilde{h}_{i,i}} x_i$, where ${\tilde{h}_{i,i}}$ is the residual interference channel coefficient after self-cancellation at node $i$, and $x_i$ ($E\{|x_i|^2\} = 1$) is the transmit signal of the node. The variance of the residual self-interference is ${P_{i}}{\left| {{\tilde{h}_{i,i}}}\right|^2}$, where ${\left| {{\tilde{h}_{i,i}}}\right|^2}$ is dependent on the specific cancellation method implemented at node $i$, {e.g.}, PS or PSAC \cite{Reference28}.
\subsection{Practical Energy Consumption}\label{Sec3:2}
The efficiency of a PA is defined to be the ratio between the desired average transmit power and the actual power consumed at the PA. In practice, PAs are non-ideal, and the PA efficiency changes (in many cases, nonlinearly) along with the output transmit power. The representative models of non-ideal PA are TPA and envelope-tracking PA~(ETPA). The energy consumptions of TPA and ETPA are given by~\cite{EEdelay}
\begin{equation}\label{TPA_ori}
    \Psi_{TPA}(P_i) \approx \frac{\sqrt{P_i P_{MAX,i} }}{\eta_{MAX,i}}, \ \ i \in \{ a, r, b \},
\end{equation}
\begin{equation}\label{ETPA_ori}
    \Psi_{ETPA}(P_i) \approx \frac{P_i + u P_{MAX,i}}{(1+u) \eta_{MAX,i}}, \ \ i \in \{ a, r, b \},
\end{equation}
where $P_i$ is the mean output transmit power of node $i$, $P_{MAX,i}$ and $\eta_{MAX,i}$ are the maximum PA output power and the maximum PA efficiency of node $i$, respectively, and $u \approx 0.0082$ is a PA dependent parameter for ETPA \cite{EEdelay}.

Let $P_{\max,i}$ denote the maximum average transmit power of node $i$, which is important to the calculation of energy consumption. Typically, $P_{\max,i}$ is around $7\sim8$ dB lower than $P_{MAX,i}$, since the peak-to-average-power ratio~(PAPR), defined by $\kappa_i = P_{MAX,i}/P_{\max,i}$, is around $7 \sim 8$ dB in modern communication systems. Also let $\eta_{\max,i}$ denote the PA efficiency associated with $P_{\max,i}$. We can rewrite (\ref{TPA_ori}) and (\ref{ETPA_ori}), w.r.t. $P_{\max,i}$, as
\begin{equation}\label{TPA}
    \Psi_{TPA}(P_i) \approx \frac{\sqrt{P_i P_{\max,i} }}{\eta_{\max,i}}, \ \ i \in \{ a, r, b \},
\end{equation}
\begin{equation}\label{ETPA}
    \Psi_{ETPA}(P_i) \approx \frac{P_i + u \kappa_i P_{\max,i}}{(1+u \kappa_i) \eta_{\max,i}}, \ \ i \in \{ a, r, b \},
\end{equation}
where, in the case of TPA, $\eta_{\max,i}=\eta_{MAX,i}/\sqrt{\kappa_i}$ is derived by setting the right-hand sides (RHSs) of (\ref{TPA_ori}) and (\ref{TPA}) equal and solving the equation for $\eta_{\max,i}$; in the case of ETPA, $\eta_{\max,i}=\frac{1+u}{1+u\kappa_i}\eta_{MAX,i}$ is derived by setting the RHSs of (\ref{ETPA_ori}) and (\ref{ETPA}) equal and solving the equation for $\eta_{\max,i}$. Clearly, the ideal PA is a special case of ETPA by letting $u = 0$.

The practical circuit power is also non-negligible and transmission-dependent. It can be decomposed into a static component $P_{base,i}$, which drives hardware; and a dynamic component $P_{c,i}$, which accounts for analog and digital signal processing. $P_{c,i}=\varepsilon R$ depends on the transmit (receive) data rate $R$, where the coefficient $\varepsilon$ is the power consumption per unit data rate~\cite{EnergyCom}. As such, the total transmit circuit power can be written as
\begin{equation}\label{ptx}
    P_{tx,i}=\Psi[P_i(R)] + \varepsilon R + P_{base,i}, \ \ i \in \{ a, r, b \}.
\end{equation}
The total receive circuit power can be written as
\begin{equation}\label{prx}
    P_{rx,i}=\varepsilon R + P_{base,i}, \ \ i \in \{ a, r, b \}.
\end{equation}
The power consumption of a node being in the idle mode (neither transmitting or receiving) is set to be constant, denoted by $P_{idle,i}$, due to its independence of $P_i$.
\subsection{TWR Transmission Strategies}\label{Sec3:3}
\begin{figure}[!t]
\centering
\subfloat[HD-TWR-2TS with PNC.]{\hspace{-4mm} \label{TransmissionModel_a} \includegraphics[scale=0.53]{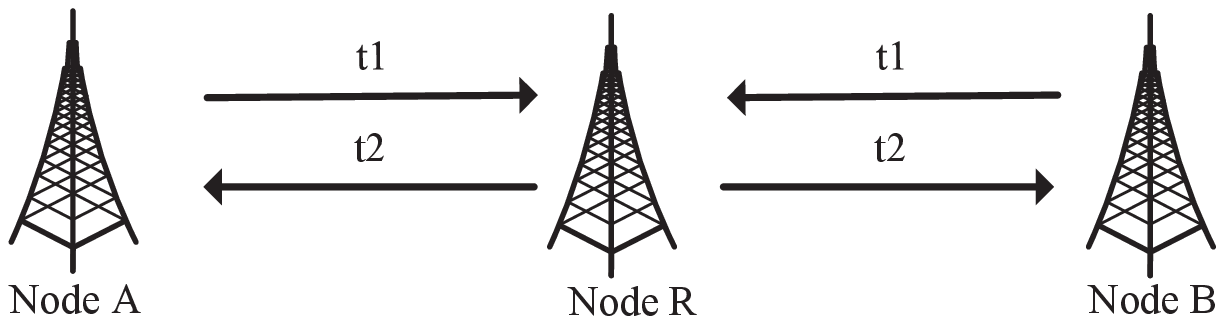}}
\vspace{-3mm}
\hfil
\centering
\subfloat[FD-TWR-2TS with FD transmission.]{\hspace{-4mm} \label{TransmissionModel_b} \includegraphics[scale=0.53]{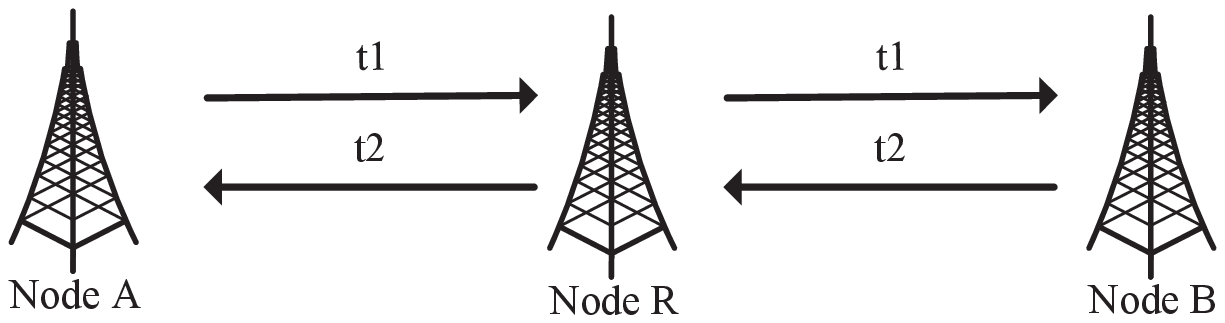}}
\vspace{-3mm}
\hfil
\centering
\subfloat[FD-TWR-1TS with FD transmission and PNC.]{\hspace{-4mm} \label{TransmissionModel_c} \includegraphics[scale=0.53]{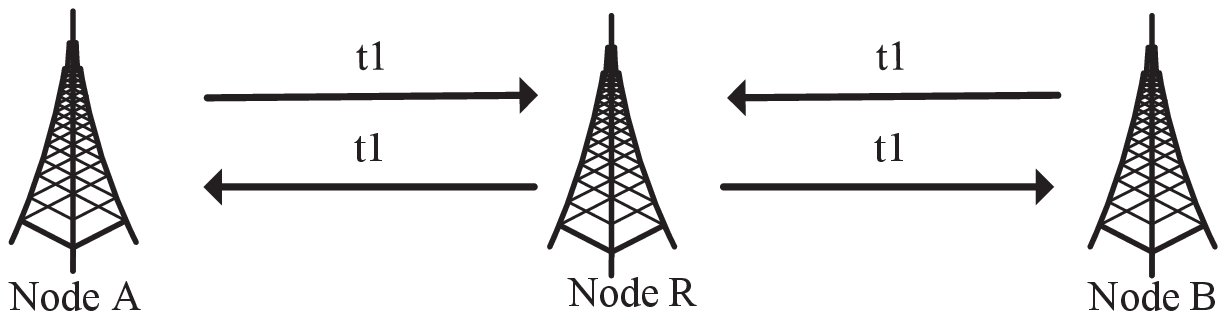}}
\caption{Illustration of the three transmission strategies.}
\label{TransmissionModel}
\end{figure}
Two strategies of FD-TWR are considered, namely, FD-TWR-1TS and FD-TWR-2TS, as shown in Fig.~\ref{TransmissionModel}. FD-TWR-2TS is relatively basic, where only the node $R$ is enabled with FD, and nodes $A$ and $B$ operate in the HD mode. As a result, two timeslots are required to complete TWR. FD-TWR-1TS is much advanced, where not only node $R$, but nodes $A$ and $B$ are also FD enabled. As a result, only a single timeslot is required to accomplish TWR.

Consider a fixed time frame of $T$ seconds, within which a round of TWR is accomplished. Let $t_k$\,($k \in \{1\}$ in FD-TWR-1TS, and $k \in \{ 1,2\}$ for FD-TWR-2TS) define the duration of the $k$-th timeslot within $T$, as shown in Fig.~\ref{TransmissionModel}, where $\sum_{\forall k} t_k\leq T$. The nodes switch to an idle mode after $\sum_{\forall k} t_k$ till the end of the frame.

For ease of understanding, we start with FD-TWR-2TS, then move to FD-TWR-1TS.
\subsubsection{FD-TWR-2TS}
In this case, two timeslots are required. Nodes $A$ and $B$, each equipped with a single antenna, transmit and receive in an alternating manner in the two timeslots, using the HD mode. Node $R$, equipped with two antennas (one for transmission and one for reception), transmits and receives simultaneously in each timeslot using FD. Specifically, in the first of the two timeslots, node $A$ transmits data (${x_a}$) to the node $R$ which receives and immediately forwards the data ($x_r^a$) to node $B$; and the other way around (${x_b}$ and $x_r^b$) in the second timeslot. The DF mode is used in FD-TWR-2TS. There can be a one-frame delay from node $R$ receiving the signals till node $R$ forwarding the signals, but this pipeline delay does not compromise the system capacity.

With the self-interference (partly) cancelled, the signal received at node $R$ in the first timeslot can be given by
\begin{equation}\label{yr1-FD2TS}
{y_r} = \sqrt {{P_a}} {h_{a,r}}{x_a} + \sqrt {{P_{r,b}}} {\tilde{h}_{r,r}}  x_r^a + {n_r}.
\end{equation}
The signal received at node $B$ in the first timeslot can be given by
\begin{equation}\label{yu-FD2TS}
{y_b} = \sqrt {{P_{r,b}}} {h_{r,b}}x_r^a + {n_b}.
\end{equation}
Likewise, the signals received at nodes $R$ and $A$ in the second timeslot can be given respectively by
\begin{equation}\label{yr2-FD2TS}
{y_r} = \sqrt {{P_b}} {h_{b,r}}{x_b} + \sqrt {{P_{r,a}}} {\tilde{h}_{r,r}}  x_r^b + {n_r},
\end{equation}
\begin{equation}\label{yb-FD2TS}
{y_a} = \sqrt {{P_{r,a}}} {h_{r,a}}x_r^b + {n_a},
\end{equation}
where ${\tilde{h}_{r,r}} = \sqrt{1/\alpha} h_{r,r}$, $E\left\{ {{{\left| {{x_a}} \right|}^2}} \right\} = 1$, $E\left\{ {{{\left| {{x_b}} \right|}^2}} \right\} = 1$, $E\left\{ {{{\left| {x_r^a} \right|}^2}} \right\} = 1$, and $E\left\{ {{{\left| {x_r^b} \right|}^2}} \right\} = 1$.

As discussed in Section~\ref{Sec3:1}, the residual self-interference at node $R$, $e_r^i = \sqrt {{P_{r,j}}} {\tilde{h}_{r,r}} x_r^i$, yields $\mathcal{CN}(0,P_{r,j}|{\tilde{h}_{r,r}}|^2)$, where $i\in\{a,b\}$ and $j\in\{b,a\}$.
This is because the transmit and receive antennas of a FD node are typically isolated, or carefully calibrated to eliminate the line-of-sight (LOS). Moreover, AC can whiten self-interference during signal processing~\cite{ResidualNoise,ResidualNoise1}. As a result, the channel between the transmitter and the receiver at a node typically follows the Rayleigh fading~\cite{Experiment-Driven}. The channel capacities $C_{a,r}$ and $C_{r,b}$ in the forward link direction in the first timeslot satisfy
\begin{equation}\label{Rbr-FD2TS}
{C_{a,r}} = \frac{{t_1}}{T}W{\log _2}(1 + \frac{{{P_a}{{\left| {{h_{a,r}}} \right|}^2}}}{{{P_{r,b}}{{\left| {{\tilde{h}_{r,r}}} \right|}^2} + \sigma _r^2}}),
\end{equation}
\begin{equation}\label{Rru-FD2TS}
{C_{r,b}} = \frac{{t_1}}{T}W{\log _2}(1 + \frac{{{P_{r,b}}{{\left| {{h_{r,b}}} \right|}^2}}}{{\sigma _b^2}}).
\end{equation}

Likewise, the capacities $C_{b,r}$ and $C_{r,a}$ in the reverse link in the second timeslot satisfy
\begin{equation}\label{Rur-FD2TS}
{C_{b,r}} = \frac{{t_2}}{T}W{\log _2}(1 + \frac{{{P_b}{{\left| {{h_{b,r}}} \right|}^2}}}{{{P_{r,a}}{{\left| {{\tilde{h}_{r,r}}} \right|}^2} + \sigma _r^2}}),
\end{equation}
\begin{equation}\label{Rrb-FD2TS}
{C_{r,a}} = \frac{{t_2}}{T}W{\log _2}(1 + \frac{{{P_{r,a}}{{\left| {{h_{r,a}}} \right|}^2}}}{{\sigma _a^2}}).
\end{equation}

Finally, the actual achievable data rate in each of the two link directions can be given by
\begin{equation}\label{AR-HD2TS}
{C_{a,b}} = \min \left\{ {{C_{a,r}},{C_{r,b}}} \right\},\ {C_{b,a}} = \min \left\{ {{C_{b,r}},{C_{r,a}}} \right\}.
\end{equation}
\subsubsection{FD-TWR-1TS}
In this case, only a single timeslot is required, as mentioned earlier. The three nodes, each equipped with a transmit antenna and a receive antenna, operate in the FD mode during the timeslot. Nodes $A$ and $B$ transmit $x_a$ and $x_b$ to node $R$, respectively, meanwhile node $R$ transmits $x_r$ to the nodes $A$ and $B$. Additionally, PNC is adopted to generate ${x_r}$ as a function of ${x_a}$ and ${x_b}$~\cite{Reference5,Reference24,ResidualNoise1}.

The received signals at the three nodes can be given by
\begin{equation}\label{yr-FD1TS}
  {y_r} = \sqrt {{P_a}} {h_{a,r}}{x_a} + \sqrt {{P_b}} {h_{b,r}}{x_b} + \sqrt {{P_r}} {{\tilde{h}_{r,r}}} {x_r} + {n_r},
\end{equation}
\begin{equation}\label{yb-FD1TS}
{y_a} = \sqrt {{P_r}} {h_{r,a}}{x_r} + \sqrt {{P_a}} {{\tilde{h}_{a,a}}} {x_a} + {n_a},
\end{equation}
\begin{equation}\label{yu-FD1TS}
{y_b} = \sqrt {{P_r}} {h_{r,b}}{x_r} + \sqrt {{P_b}} {{\tilde{h}_{b,b}}} {x_b} + {n_b},
\end{equation}
where $E\left\{ {{x_i}} \right\} = 0$; $E\left\{ {{{\left| {{x_i}} \right|}^2}} \right\} = 1\ (i \in \left\{ {a,b,r} \right\})$; $\sqrt {{P_i}} {{\tilde{h}_{i,i}}} {x_i}$ is the residual self-interference at node $i\in \{a, b, r\}$; and $e_i = \sqrt {{P_i}} {\tilde{h}_{i,i}} x_i$ yields $\mathcal{CN}(0,{P_i}{\left| {{\tilde{h}_{i,i}}}\right|^2})$.

We consider nested lattice code and structured binning, as defined by three $n$-dimensional lattices $\Lambda_C^n$, $\Lambda_1^n$ and $\Lambda_2^n$. Assume $\Lambda_1^n \subseteq \Lambda_2^n \subseteq \Lambda_C^n$. $\Lambda_C^n$ is used as codewords, and the Voronoi regions of $\Lambda_1^n$ and $\Lambda_2^n$ are the shaping regions of nodes $A$ and $B$. Node $R$ also uses $\Lambda_1^n$ as its shaping region. Rather than recovering the exact signals of nodes $A$ and $B$, node $R$ recovers a structured binned version of the signals combined,  by evaluating its Euclidean distance to the codewords~\cite{latticecode}, exploiting maximum-likelihood estimation~\cite{latticecodeOrin1,latticecodeOrin2}, or minimum mean square error (MMSE)~\cite{latticecodecapacity}. On the receipt of the binned signals, nodes $A$ and $B$ can recover the desired signal from each other by canceling the instantaneous self-interference and subtracting their own signals from the binned signals~\cite{latticecode}.

There can be a one-frame delay from node $R$ receiving the signals till it forwarding the binned version of the signals. In other words, the instantaneous self-interference is independent of the signals that nodes $A$ and $B$ subtract from their received binned signals, as described in~\cite{latticecode}. Such a delay is a pipeline delay, and does not compromise the throughput or EE of the network under stable channel conditions. This paper focuses on the allocation of temporal and energy resources to maximize the EE in FD-TWR systems using nested lattice coding. Designing nested lattice codes is beyond the scope of the paper.

Considering the nested lattice coding with structured binning~\cite{latticecode}, the channel capacities from nodes $A$ and $B$ to node $R$ can be given by
\begin{equation}\label{Rbr-FD1TS}
{C_{a,r}} \hspace{-1mm} = \hspace{-1mm} \frac{t_1}{T}W{\log _2}(\frac{{{P_a}{{\left| {{h_{a,r}}} \right|}^2}}}{{{P_a}{{\left| {{h_{a,r}}} \right|}^2} \hspace{-1mm} + \hspace{-1mm} {P_b}{{\left| {{h_{b,r}}} \right|}^2}}} \hspace{-1mm} + \hspace{-1mm} \frac{{{P_a}{{\left| {{h_{a,r}}} \right|}^2}}}{{{P_r}{{\left| {{\tilde{h}_{r,r}}} \right|}^2} \hspace{-1mm} + \hspace{-1mm} \sigma _r^2}}),
\end{equation}
\begin{equation}\label{Rur-FD1TS}
{C_{b,r}} \hspace{-1mm} = \hspace{-1mm} \frac{t_1}{T}W{\log _2}(\frac{{{P_b}{{\left| {{h_{b,r}}} \right|}^2}}}{{{P_a}{{\left| {{h_{a,r}}} \right|}^2} \hspace{-1mm} + \hspace{-1mm} {P_b}{{\left| {{h_{b,r}}} \right|}^2}}} \hspace{-1mm} + \hspace{-1mm} \frac{{{P_b}{{\left| {{h_{b,r}}} \right|}^2}}}{{{P_r}{{\left| {{\tilde{h}_{r,r}}} \right|}^2} \hspace{-1mm} + \hspace{-1mm} \sigma _r^2}}).
\end{equation}

The data rates from node $R$ to nodes $A$ and $B$ can be respectively given by
\begin{equation}\label{Rrb-FD1TS}
{C_{r,a}} = \frac{t_1}{T}W{\log _2}(1 + \frac{{{P_r}{{\left| {{h_{r,a}}} \right|}^2}}}{{{P_a}{{\left| {{\tilde{h}_{a,a}}} \right|}^2} + \sigma _a^2}}),
\end{equation}
\begin{equation}\label{Rru-FD1TS}
{C_{r,b}} = \frac{t_1}{T}W{\log _2}(1 + \frac{{{P_r}{{\left| {{h_{r,b}}} \right|}^2}}}{{{P_b}{{\left| {{\tilde{h}_{b,b}}} \right|}^2} + \sigma _b^2}}),
\end{equation}
where $\left(P_i {{\left| {{\tilde{h}_{i,i}}} \right|}^2} + \sigma _i^2\right)$ is the variance of ($e_i + n_i$), due to the fact that $e_i$ and $n_i$ are independent.

Substituting (\ref{Rbr-FD1TS})-(\ref{Rru-FD1TS}) into (\ref{AR-HD2TS}), we can finally obtain the achievable data rate in both link directions.
\section{EE Maximization of FD-TWR-2TS}\label{Sec4}
In this section, we maximize the EE of FD-TWR-2TS, given the data rate requirements $R_{rl}$ and $R_{fl}$, under non-ideal PAs and non-negligible circuit power. Here, $R_{rl}$ and $R_{fl}$ are the data rate requirements in the reverse link ({i.e.}, from node $B$ to node $A$) and the forward link ({i.e.}, from node $A$ to node $B$), respectively.

The EE is defined as the ratio between the average total data rate in both link directions and the average total power consumed by the nodes~\cite{EEmetric}, as given by
\begin{equation}\label{EE_indicator}
{\eta _E} = \frac{{R_{rl}} + {R_{fl}}}{E_{{\rm total}}/T} = \frac{{({R_{rl}} + {R_{fl}})T}}{{{E_{{\rm total}}}}},
\end{equation}
where the second equality indicates that the EE is equivalent to the ratio between the number of bits to be transmitted (in both directions) within $T$ and the total energy required to transmit the bits, denoted by $E_{{\rm total}}$. To this end, given the number of bits to be sent within $T$, {i.e.}, ($R_{rl} + R_{fl})T$, maximizing $\eta_E$ is equivalent to minimizing $E_{{\rm total}}$.

Minimizing $E_{{\rm total}}$ facilitates maximizing the EE, under non-negligible circuit power. The reason is that, under non-negligible circuit power, maximizing the EE may require the nodes to transmit for part of a frame to leverage the non-negligible signal processing energy and circuit standby energy. The optimal transmit rate may switch to null during a frame. The direct maximization of the EE, {i.e.}, directly maximizing $\eta_E$, would be unsuitable, due to such change of the data rate.

Therefore, the EE maximization can be formulated as
\begin{equation}
\mathop {\min }\limits_{{P_a},{P_b},{P_{r,a}},{P_{r,b}},{t_1},{t_2}}\ \ {E_{{\rm total}}} \tag{\textbf{P1}} \label{EE_Opt_FD_2TS}
\end{equation}
\begin{equation}\nonumber
{\rm s.t.}\ \ \min \left\{ {{C_{b,r}},{C_{r,a}}} \right\} \ge {R_{rl}},\ \min \left\{ {{C_{a,r}},{C_{r,b}}} \right\} \ge {R_{fl}};
\end{equation}
\begin{equation}\nonumber
0 \le \Psi \left( {{P_i}} \right) \le {P_{\max ,i}},\ i \in \{a,b,r\};
\end{equation}
\begin{equation}\nonumber
{t_1} + {t_2} \le T,\ 0 \le {t_1} \le T,\ 0 \le {t_2} \le T,
\end{equation}
where the total energy consumption ${E_{{\rm total}}} = ( {P_{tx,a}} + {P_{tx,r}} + {P_{rx,b}} + {P_{rx,r}} ){t_1} + ({P_{tx,b}} + {P_{tx,r}} + {P_{rx,a}} + {P_{rx,r}}){t_2} + P_{idle}(T - {t_1} - {t_2})$ and $P_{idle} = {P_{idle,a}} + {P_{idle,b}} + {P_{idle,r}}$.

(\ref{EE_Opt_FD_2TS}) is non-convex w.r.t. the transmit powers and durations, {i.e.}, $P_i$ ($i\in \{a, b, r\}$) and $t_j$ ($j\in \{1, 2\}$). This is because that under both TPA and ETPA, the logarithmic data constraint is non-convex in $P_i$ and $t_j$, due to the non-convexity nature of logarithm. This leads to a non-convex feasible set for the EE maximization problem of interest. Further, under TPA, the objective of (\ref{EE_Opt_FD_2TS}) is non-convex, as can be rigorously verified through the Hessian matrix $\mathbf{H}$, as given by
\begin{equation}\nonumber
\mathbf{H} \hspace{-1mm} = \hspace{-1mm} \left[ \hspace{-1mm} {\begin{array}{*{20}{c}}
-\Gamma_1(a) t_1 \hspace{-1mm} & \hspace{-1mm} 0 \hspace{-1mm} & \hspace{-1mm} 0 \hspace{-1mm} & \hspace{-1mm} \Gamma_2(a) \hspace{-1mm} & \hspace{-1mm} 0\\
0 \hspace{-1mm} & \hspace{-1mm} -\Gamma_1(r)(t_1 \hspace{-1mm} + \hspace{-1mm} t_2) \hspace{-1mm} & \hspace{-1mm} 0 \hspace{-1mm} & \hspace{-1mm} \Gamma_2(r) \hspace{-1mm} & \hspace{-1mm} \Gamma_2(r)\\
0 \hspace{-1mm} & \hspace{-1mm} 0 \hspace{-1mm} & \hspace{-1mm} -\Gamma_1(b) t_2 \hspace{-1mm} & \hspace{-1mm} 0 \hspace{-1mm} & \hspace{-1mm} \Gamma_2(b)\\
\Gamma_2(a) \hspace{-1mm} & \hspace{-1mm} \Gamma_2(r) \hspace{-1mm} & \hspace{-1mm} 0 \hspace{-1mm} & \hspace{-1mm} 0 \hspace{-1mm} & \hspace{-1mm} 0\\
0 \hspace{-1mm} & \hspace{-1mm} \Gamma_2(r) \hspace{-1mm} & \hspace{-1mm} \Gamma_2(b) \hspace{-1mm} & \hspace{-1mm} 0 \hspace{-1mm} & \hspace{-1mm} 0
\end{array}} \hspace{-1mm} \right]
\end{equation}
where
\begin{equation}\nonumber
\Gamma_1(i)=\frac{P^2_{\max,i}}{4 \eta_{\max,i} {(P_i P_{\max,i})}^{2/3}}>0,\ i \in \{a,r,b\},
\end{equation}
\begin{equation}\nonumber
\Gamma_2(i)=\frac{P_{\max,i}}{2\eta_{\max,i}\sqrt{P_i P_{\max,i}}}>0,\ i \in \{a,r,b\}.
\end{equation}
Clearly, $\mathbf{H}$ is not positive definite, given negative leading principal minors, e.g., the first and third leading principal minors.

To convexify (\ref{EE_Opt_FD_2TS}), we can prove that the necessary condition of the optimal solution for (\ref{EE_Opt_FD_2TS}) can be given by
\begin{equation}\label{Equal_Con_FD_2TS}
{C_{b,r}} = {C_{r,a}} = {R_{rl}},\ {C_{a,r}} = {C_{r,b}} = {R_{fl}}.
\end{equation}
The detailed proof is provided in Appendix~\ref{Apx1}.

As a result, the optimal transmit powers, $P_a^*$, $P_b^*$, $P_{r,b}^*$ and $P_{r,a}^*$, can be rewritten as the functions of the optimal transmit durations, $t_1^*$ and $t_2^*$, as given by
\begin{equation}\label{Pru_FD_2TS}
{P_{r,b}^*}\left( {{t_1^*}} \right) = \frac{{\sigma _b^2}}{{{{\left| {{h_{r,b}}} \right|}^2}}}\left( {{2^{\frac{{{R_{fl}}T}}{{{t_1^*}W}}}} - 1} \right);
\end{equation}
\begin{equation}\label{Prb_FD_2TS}
{P_{r,a}^*}\left( {{t_2^*}} \right) = \frac{{\sigma _a^2}}{{{{\left| {{h_{r,a}}} \right|}^2}}}\left( {{2^{\frac{{{R_{rl}}T}}{{{t_2^*}W}}}} - 1} \right);
\end{equation}
\begin{equation}\label{Pb_FD_2TS}
{P_a^*}\left( {{t_1^*}} \right) \hspace{-1mm} = \hspace{-1mm} \frac{{\sigma _r^2}}{{{{\left| {{h_{a,r}}} \right|}^2}}}\left( {{2^{\frac{{{R_{fl}}T}}{{{t_1^*}W}}}} \hspace{-1mm} - \hspace{-1mm} 1} \right) \hspace{-1mm} + \hspace{-1mm} \frac{{\sigma _b^2{{\left| {{\tilde{h}_{r,r}}} \right|}^2}}}{{{{\left| {{h_{a,r}}} \right|}^2}{{\left| {{h_{r,b}}} \right|}^2}}}{\left( {{2^{\frac{{{R_{fl}}T}}{{{t_1^*}W}}}} \hspace{-1mm} - \hspace{-1mm} 1} \right)^2};
\end{equation}
\begin{equation}\label{Pu_FD_2TS}
{P_b^*}\left( {{t_2^*}} \right) \hspace{-1mm} = \hspace{-1mm} \frac{{\sigma _r^2}}{{{{\left| {{h_{b,r}}} \right|}^2}}}\left( {{2^{\frac{{{R_{rl}}T}}{{{t_2^*}W}}}} \hspace{-1mm} - \hspace{-1mm} 1} \right) \hspace{-1mm} + \hspace{-1mm} \frac{{\sigma _a^2{{\left| {{\tilde{h}_{r,r}}} \right|}^2}}}{{{{\left| {{h_{b,r}}} \right|}^2}{{\left| {{h_{r,a}}} \right|}^2}}}{\left( {{2^{\frac{{{R_{rl}}T}}{{{t_2^*}W}}}} \hspace{-1mm} - \hspace{-1mm} 1} \right)^2}.
\end{equation}

Therefore, (\ref{EE_Opt_FD_2TS}) can be reformulated to only optimize the transmit durations, $t_i$ ($i=1,2$), as given by
\begin{align}
\mathop {\min }\limits_{\mathbf{t}} \ \ &{E_{{\rm total}}} = E\left( {\mathbf{t}} \right) \label{EE_Opt_General}\\
{\rm s.t.}\ \ &{{\mathbf{t}}_{\min }} \le \mathbf{t}; \nonumber \\
&{\mathbf{1}^T}{\mathbf{t}} \le T, \nonumber
\end{align}
where $E(\cdot)$ is the objective function denoting the overall energy consumption. ${\mathbf{t}}$ is a column vector collecting both the time durations to be optimized, ${\mathbf{t}}_{min}$ is a column vector collecting the minimum value of each time duration $t_{\min,i}$, and $\mathbf{1}$ is the all-one column vector. The minimum time durations are specified by $t_i \geq t_{\min ,i}$, which can be readily calculated.

In the case of FD-TWR-2TS, $\mathbf{t}_{\min}=[t_{\min,1},t_{\min,2}]^T$. $t_{\min,1}$ and $t_{\min,2}$ are obtained by separately setting \eqref{Pru_FD_2TS}--\eqref{Pu_FD_2TS} to be no larger than the corresponding maximum transmit powers, solving the inequalities, and choosing the intersection of the results.

For notation simplicity, we denote ${\lambda _{fl}} = \frac{{R_{fl}}T}{W{t_1}}$ and ${\lambda _{rl}} = \frac{{R_{rl}}T}{W{t_2}}$. $\lambda_{fl}$ and $\lambda_{rl}$ are the spectrum efficiency in the forward and reverse link directions, respectively. The following theorems dictate that \eqref{EE_Opt_General} yields a convex structure under non-ideal PAs and non-negligible circuit power.

\vspace{3mm}

\newtheorem{theorem}{\textbf{Theorem}}
\begin{theorem} \label{thtpa_FD_2TS}
\textit{Under TPA, (\ref{EE_Opt_General}) is convex for FD-TWR-2TS.}
\end{theorem}

\vspace{3mm}

\begin{IEEEproof}
The proof starts by substituting (\ref{Pru_FD_2TS})--(\ref{Pu_FD_2TS}) into the objective function of (\ref{EE_Opt_FD_2TS}). The objective function can be rewritten as
\begin{align}
&E\left( {{t_1},{t_2}} \right) =  \nonumber \\
& \left\{ {\sqrt {\beta_{1,1} \hspace{-1mm} \times \hspace{-1mm} {2^{\frac{{{R_{fl}}T}}{{{t_1}W}}}} \hspace{-1mm} + \hspace{-0.5mm} \gamma_{1} \hspace{-1mm} \times \hspace{-1mm} {{\left( {{2^{\frac{{{R_{fl}}T}}{{{t_1}W}}}}} \right)}^2}} \hspace{-1mm} + \hspace{-1mm} \sqrt {\alpha_{1,1} \hspace{-1mm} \times \hspace{-1mm} {2^{\frac{{{R_{fl}}T}}{{{t_1}W}}}}} } \right\}{t_{1}} + \nonumber \\
&\left\{ {\sqrt {\beta_{1,2} \hspace{-1mm} \times \hspace{-1mm} {2^{\frac{{{R_{rl}}T}}{{{t_2}W}}}} \hspace{-1mm} + \hspace{-0.5mm} \gamma_{2} \hspace{-1mm} \times \hspace{-1mm} {{\left( {{2^{\frac{{{R_{rl}}T}}{{{t_2}W}}}}} \right)}^2}} \hspace{-1mm} + \hspace{-1mm} \sqrt {\alpha_{1,2} \hspace{-1mm} \times \hspace{-1mm} {2^{\frac{{{R_{rl}}T}}{{{t_2}W}}}}} } \right\}{t_2} + \nonumber \\
&{P_{1,1}}{t_1} + {P_{1,2}}{t_2} + ({P_{idle,a}} + {P_{idle,b}} + {P_{idle,r}})T, \nonumber
\end{align}
\begin{equation}\nonumber
\alpha_{1,1} = \frac{{{P_{\max ,r}}\sigma _b^2}}{{{\eta _{\max ,r}^2}{{\left| {{h_{r,b}}} \right|}^2}}},\ \alpha_{1,2} = \frac{{{P_{\max ,r}}\sigma _a^2}}{{{\eta _{\max ,r}^2}{{\left| {{h_{r,a}}} \right|}^2}}},
\end{equation}
\begin{equation}\nonumber
\beta_{1,1} = \frac{{{P_{\max ,a}}\sigma _r^2}}{{{\eta _{\max ,a}^2}{{\left| {{h_{a,r}}} \right|}^2}}},\ \beta_{1,2} = \frac{{{P_{\max ,b}}\sigma _r^2}}{{{\eta _{\max ,b}^2}{{\left| {{h_{b,r}}} \right|}^2}}},
\end{equation}
\begin{equation}\nonumber
\gamma_{1} = \frac{{{P_{\max ,a}}\sigma _b^2{{\left| {{\tilde{h}_{r,r}}} \right|}^2}}}{{{\eta _{\max ,a}^2}{{\left| {{h_{a,r}}} \right|}^2}{{\left| {{h_{r,b}}} \right|}^2}}},\ \gamma_{2} = \frac{{{P_{\max ,b}}\sigma _a^2{{\left| {{\tilde{h}_{r,r}}} \right|}^2}}}{{{\eta _{\max ,b}^2}{{\left| {{h_{b,r}}} \right|}^2}{{\left| {{h_{r,a}}} \right|}^2}}},
\end{equation}
where
\begin{equation}\nonumber
{P_{1,1}} \hspace{-1mm} = \hspace{-1mm} {P_{base,a}} \hspace{-1mm} + \hspace{-1mm} {P_{base,b}} \hspace{-1mm} + \hspace{-1mm} 2{P_{base,r}} \hspace{-1mm} + \hspace{-1mm} 4 \varepsilon R_{fl} \hspace{-1mm} - \hspace{-1mm} {P_{idle,a}} \hspace{-1mm} - \hspace{-1mm} {P_{idle,b}} \hspace{-1mm} - \hspace{-1mm} {P_{idle,r}},
\end{equation}
\begin{equation}\nonumber
{P_{1,2}} \hspace{-1mm} = \hspace{-1mm} {P_{base,a}} \hspace{-1mm} + \hspace{-1mm} {P_{base,b}} \hspace{-1mm} + \hspace{-1mm} 2 {P_{base,r}} \hspace{-1mm} + \hspace{-1mm} 4 \varepsilon R_{rl} \hspace{-1mm} - \hspace{-1mm} {P_{idle,a}} \hspace{-1mm} - \hspace{-1mm} {P_{idle,b}} \hspace{-1mm} - \hspace{-1mm} {P_{idle,r}}.
\end{equation}

$E\left( {{t_1},{t_2}} \right)$ is convex, because Hessian matrix is positive definite, as given by
\[\left[ {\begin{array}{*{20}{c}}
A&0\\
0&B
\end{array}} \right]\]
\begin{align}
A = &\frac{1}{{{4\alpha^2_{1,1}}{q_1}^{\frac{3}{2}}{t_1}^3{W^2}}} [{\beta^2_{1,1}} \hspace{-1mm} \sqrt {\alpha_{1,1}2^{\lambda _{fl}}} \hspace{-1mm} + \hspace{-1mm} \gamma_{1}2^{\lambda _{fl}}( {4p_1 \hspace{-1mm} + \hspace{-1mm} \alpha_{1,1}\sqrt {q_1} } ) \nonumber \\
&\hspace{-1mm} + \beta_{1,1}( {6p_1 \hspace{-1mm} + \hspace{-1mm} \alpha_{1,1}\sqrt {q_1} } ) ] ({\alpha_{1,1}2^{\lambda _{fl}}})^{\frac{3}{2}}R_{fl}^2{T^2}\log (2)^2 \hspace{-1mm} > \hspace{-1mm} 0, \nonumber
\end{align}
\begin{align}
B = &\frac{1}{{4{\alpha^2_{1,2}}{q_2}^{\frac{3}{2}}{t_2}^3{W^2}}} [ {\beta^2_{1,2}} \hspace{-1mm} \sqrt {\alpha_{1,2}2^{\lambda _{rl}}} \hspace{-1mm} + \hspace{-1mm} \gamma_{2}2^{\lambda _{rl}}(4p_2 \hspace{-1mm} + \hspace{-1mm} \alpha_{1,2}\sqrt {q_2}) \nonumber \\
&\hspace{-1mm} + \beta_{1,2}(6p_2 \hspace{-1mm} + \hspace{-1mm} \alpha_{1,2}\sqrt {q_2}) ] (\alpha_{1,2}2^{\lambda _{rl}})^\frac{3}{2}R_{rl}^2{T^2}\log (2)^2 \hspace{-1mm} > \hspace{-1mm} 0, \nonumber
\end{align}
where $p_1 = \frac{{{{\left( {\alpha_{1,1}2^{\lambda _{fl}}} \right)}^{\frac{3}{2}}}\gamma_{1}}}{{\alpha_{1,1}}},\ q_1 = 2^{\lambda _{fl}}\left( {\beta_{1,1} + \gamma_{1}2^{\lambda _{fl}}} \right),\ p_2 = \frac{{{{\left( {\alpha_{1,2}2^{\lambda _{rl}}} \right)}^{\frac{3}{2}}}\gamma_{2}}}{{\alpha_{1,2}}},\ q_2 = 2^{\lambda _{rl}}\left( {\beta_{1,2} + \gamma_{2}2^{\lambda _{rl}}} \right)$
\end{IEEEproof}

\vspace{3mm}

\begin{theorem} \label{thetpa_FD_2TS}
\textit{Under ETPA, (\ref{EE_Opt_General}) is convex for FD-TWR-2TS.}
\end{theorem}

\vspace{3mm}

\begin{IEEEproof}
To prove this, we substitute (\ref{Pru_FD_2TS})--(\ref{Pu_FD_2TS}) into the objective function of (\ref{EE_Opt_FD_2TS}). The objective function can be written as
\begin{align}
E\left( {{t_1},{t_2}} \right) & = \left\{ {\alpha_{2,1}\left( {{2^{\frac{{{R_{fl}}T}}{{{t_1}W}}}} - 1} \right) + \beta_{2,1}{{\left( {{2^{\frac{{{R_{fl}}T}}{{{t_1}W}}}} - 1} \right)}^2}} \right\}{t_1} \nonumber \\
&+ \left\{ {\alpha_{2,2}\left( {{2^{\frac{{{R_{rl}}T}}{{{t_2}W}}}} - 1} \right) + \beta_{2,2}{{\left( {{2^{\frac{{{R_{rl}}T}}{{{t_2}W}}}} - 1} \right)}^2}} \right\}{t_2} \nonumber \\
&+ {P_{2,1}}{t_1} + {P_{2,2}}{t_2} + ({P_{idle,a}} + {P_{idle,b}} + {P_{idle,r}})T, \nonumber
\end{align}
\begin{equation}\nonumber
\alpha_{2,1} = \frac{{\sigma _r^2}}{{\left( {1 + u \kappa_a} \right){\eta _{\max ,a}}{{\left| {{h_{a,r}}} \right|}^2}}} + \frac{{\sigma _b^2}}{{\left( {1 + u \kappa_r } \right){\eta _{\max ,r}}{{\left| {{h_{r,b}}} \right|}^2}}},
\end{equation}
\begin{equation}\nonumber
\alpha_{2,2} = \frac{{\sigma _r^2}}{{\left( {1 + u \kappa_b} \right){\eta _{\max ,b}}{{\left| {{h_{b,r}}} \right|}^2}}} + \frac{{\sigma _a^2}}{{\left( {1 + u \kappa_r} \right){\eta _{\max ,r}}{{\left| {{h_{r,a}}} \right|}^2}}},
\end{equation}
\begin{equation}\nonumber
\beta_{2,1} = \frac{{\sigma _b^2{{\left| {{\tilde{h}_{r,r}}} \right|}^2}}}{{\left( {1 + u \kappa_a} \right){\eta _{\max ,a }}{{\left| {{h_{a,r}}} \right|}^2}{{\left| {{h_{r,b}}} \right|}^2}}},
\end{equation}
\begin{equation}\nonumber
\beta_{2,2} = \frac{{\sigma _a^2{{\left| {{\tilde{h}_{r,r}}} \right|}^2}}}{{\left( {1 + u \kappa_b} \right){\eta _{\max ,b }}{{\left| {{h_{b,r}}} \right|}^2}{{\left| {{h_{r,a}}} \right|}^2}}},
\end{equation}
where
\begin{align}\nonumber
P_{2,1} &= \frac{u \kappa_a P_{\max ,a}}{\left( {1 \hspace{-1mm} + \hspace{-1mm} u \kappa_a} \right){\eta _{\max ,a}}} \hspace{-1mm} + \hspace{-1mm} \frac{{u \kappa_r {P_{\max ,r}}}}{{\left( {1 \hspace{-1mm} + \hspace{-1mm} u \kappa_r} \right){\eta _{\max ,r}}}} \hspace{-1mm} + \hspace{-1mm} P_{base,a} \hspace{-1mm} + \hspace{-1mm} P_{base,b} \\
&+ 2 P_{base,r} + 4 \varepsilon R_{fl} - P_{idle,a} - P_{idle,b} - P_{idle,r}, \nonumber
\end{align}
\begin{align}\nonumber
P_{2,2} &= \frac{u \kappa_b P_{\max ,b}}{{\left( {1 \hspace{-1mm} + \hspace{-1mm} u \kappa_b} \right){\eta _{\max ,b}}}}\hspace{-1mm}  + \hspace{-1mm} \frac{{u \kappa_r{P_{\max ,r}}}}{{\left( {1 \hspace{-1mm} + \hspace{-1mm} u \kappa_r} \right){\eta _{\max ,r}}}} \hspace{-1mm} + \hspace{-1mm} P_{base,a} \hspace{-1mm} + \hspace{-1mm} P_{base,b} \\
&+ 2 P_{base,r} + 4 \varepsilon R_{rl} - P_{idle,a} - P_{idle,b} - P_{idle,r}. \nonumber
\end{align}

Similarly, the Hessian matrix of $E\left( {{t_1},{t_2}} \right)$ can be proved to be positive definite. As a result, $E\left( {{t_1},{t_2}} \right)$ is convex.
\end{IEEEproof}
\section{EE Maximization of FD-TWR-1TS}\label{Sec5}
In this section, we consider FD-TWR-1TS under non-ideal PAs and non-negligible circuit powers. The EE maximization of FD-TWR-1TS can be formulated as
\begin{equation}
\mathop {\min }\limits_{{P_a},{P_b},{P_r},t_1}\ \ \ {E_{{\rm total}}} \tag{\textbf{P2}} \label{EE_Opt_FD_1TS}
\end{equation}
\begin{equation}\nonumber
{\rm s.t.}\ \ \min \left\{ {{C_{b,r}},{C_{r,a}}} \right\} \ge {R_{rl}},\ \min \left\{ {{C_{a,r}},{C_{r,b}}} \right\} \ge {R_{fl}};
\end{equation}
\begin{equation}\nonumber
0 \le \Psi ({P_i}) \le {P_{\max ,i}},\ i \in \{a,b,r\};
\end{equation}
\begin{equation}\nonumber
0 \le t_1 \le T,
\end{equation}
where the total energy consumption is ${E_{{\rm total}}} = ({P_{tx,a}} + {P_{tx,b}} + {P_{tx,r}} + {P_{rx,a}} + {P_{rx,b}} + {P_{rx,r}}){t_1} + {P_{idle}}(T - {t_1})$ and ${P_{idle}} = {P_{idle,a}} + {P_{idle,b}} + {P_{idle,r}}$.

(\ref{EE_Opt_FD_1TS}) is not convex due to the non-convex feasible region posed by non-convex logarithmic data rate constraints, as well as the non-linear transmit power constraints as discussed for FD-TWR-2TS in Section~III. As proved in Appendix~\ref{Apx2}, the necessary conditions of the optimal solution for (\ref{EE_Opt_FD_1TS}) can be given by
\begin{equation}\label{Equal_RCon1}
{C_{b,r}} = {R_{rl}},{C_{a,r}} = {R_{fl}};
\end{equation}
\begin{equation}\label{Equal_RCon2}
\min \left\{ {{C_{r,a}} - {R_{rl}},{C_{r,b}} - {R_{fl}}} \right\} = 0.
\end{equation}

\begin{figure}[!t]
\centering
\hspace{-4mm}
\includegraphics[scale=0.28]{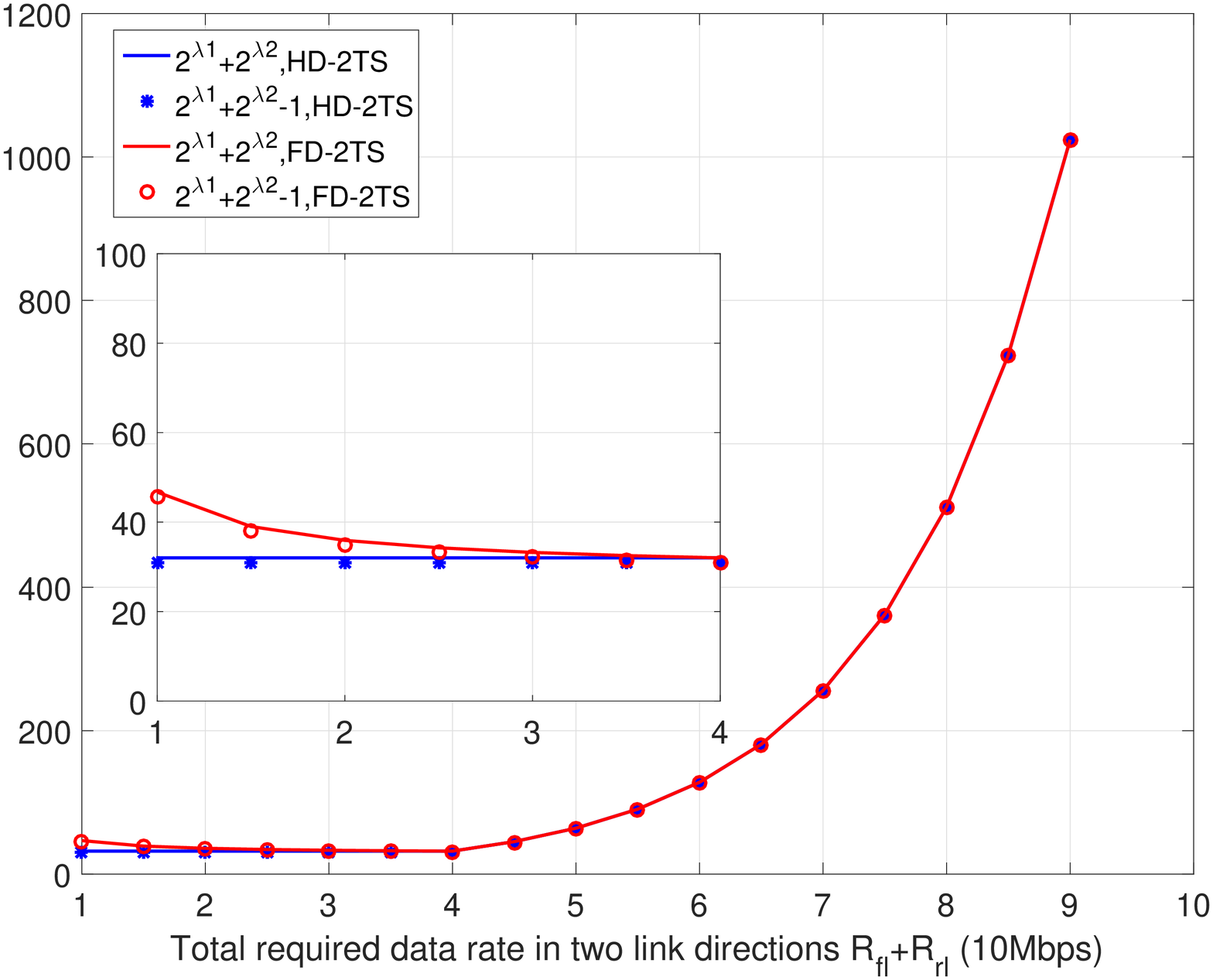}
\vspace{-3mm}
\caption{Example for the approximate relationship justification where the simulation parameters can be found in Table~I (${\lambda{1}}={\lambda _{fl}}$ and ${\lambda{2}}={\lambda _{rl}}$).}
\label{fig_approximate}
\end{figure}

We are particularly interested in the asymptotic case where the data rate requirements are high, i.e., $2^{\lambda _{fl}}+2^{\lambda _{rl}} \gg 1$. This consideration is reasonable, since FD is typically exploited to increase data rates. In the case where the data rate requirements are not high, this approximation of $2^{\lambda_{fl}}+2^{\lambda_{rl}}\gg 1$ can also hold to a great extent. This is because the transmit time of the nodes, i.e., $t_1$, is reduced to decrease circuit energy consumption and increase the EE. As a result, $\frac{T}{t_1}$ increases with the decrease of $\frac{R_{fl}}{W}$ and $\frac{R_{rl}}{W}$, avoiding significant decreases of $\lambda_{fl}=\frac{R_{fl}}{W}\frac{T}{t_1}$ and $\lambda_{rl}=\frac{R_{rl}}{W}\frac{T}{t_1}$. This approximation is reliable over a wide range of $R_{fl}$ and $R_{rl}$, as shown in Fig.~\ref{fig_approximate}.

\vspace{1mm}

In this case, solving (\ref{Rbr-FD1TS}), (\ref{Rur-FD1TS}) and (\ref{Equal_RCon1}), we can write the optimal transmit powers, $P_a$ and $P_b$, as the functions of $t_1$ and $P_r$, as given by
\begin{align}\label{Pb_FD_1TS}
{P_a}\left( {{t_1}} \right) &=  2^{\lambda _{fl}}\frac{(2^{\lambda _{fl}}+2^{\lambda _{rl}}-1)({{P_r(t_1)}{{\left| {{\tilde{h}_{r,r}}} \right|}^2} + \sigma _r^2})}{(2^{\lambda _{fl}}+2^{\lambda _{rl}}){\left| {{h_{a,r}}} \right|}^2} \nonumber \\
&\approx 2^{\lambda _{fl}}\frac{{{P_r(t_1)}{{\left| {{\tilde{h}_{r,r}}} \right|}^2} + \sigma _r^2}}{{{{\left| {{h_{a,r}}} \right|}^2}}},
\end{align}
\begin{align}\label{Pu_FD_1TS}
{P_b}\left( {{t_1}} \right) &= 2^{\lambda _{rl}}\frac{(2^{\lambda _{fl}}+2^{\lambda _{rl}}-1)({{P_r(t_1)}{{\left| {{\tilde{h}_{r,r}}} \right|}^2} + \sigma _r^2})}{(2^{\lambda _{fl}}+2^{\lambda _{rl}}){\left| {{h_{b,r}}} \right|}^2} \nonumber \\
&\approx 2^{\lambda _{rl}}\frac{{{P_r(t_1)}{{\left| {{\tilde{h}_{r,r}}} \right|}^2} + \sigma _r^2}}{{{{\left| {{h_{b,r}}} \right|}^2}}}.
\end{align}

From (\ref{Equal_RCon2}), it can be seen that $P_r$ is the larger of ${P_{r,rl}}$ and ${P_{r,fl}}$, which is the solution for the equality of (\ref{Equal_RCon2}) with (\ref{Rrb-FD1TS}) and (\ref{Rru-FD1TS}), respectively. Therefore,
\begin{equation}\label{Pr_FD_1TS}
{P_r}\left( {{t_1}} \right) = \max \{ {P_{r,rl}};{P_{r,fl}} \}.
\end{equation}

We proceed with two cases to derive the explicit transmit powers of all the three nodes, {i.e.}, nodes $R$, $A$, and $B$. First, $P_r(t_1)$ is derived, and then substituted to (\ref{Pb_FD_1TS}) and (\ref{Pu_FD_1TS}) for $P_a(t_1)$ and $P_b(t_1)$.

\textit{CASE I}: In the case of ${P_{r,rl}} \geq {P_{r,fl}}$, (\ref{Rrb-FD1TS}) meets the equality for the optimal solution. $P_r$ is solved through the equalities of (\ref{Rrb-FD1TS}) and (\ref{Pb_FD_1TS}), as given by
\begin{equation}\label{Pr1_FD_1TS}
{P_{r,1}}\left( {{t_1}} \right) \hspace{-1mm} = \hspace{-1mm} P_{r,rl} \hspace{-1mm} \approx \hspace{-1mm} \frac{{2^{\lambda _{rl}}\left( {2^{\lambda _{fl}}{{\left| {{\tilde{h}_{a,a}}} \right|}^2}\sigma _r^2 \hspace{-1mm} + \hspace{-1mm} {{\left| {{h_{a,r}}} \right|}^2}\sigma _a^2} \right)}}{{{{\left| {{h_{r,a}}} \right|}^2}{{\left| {{h_{a,r}}} \right|}^2} \hspace{-1mm} - \hspace{-1mm} 2^{\lambda _{fl}}2^{\lambda _{rl}}{{\left| {{\tilde{h}_{a,a}}} \right|}^2}{{\left| {{\tilde{h}_{r,r}}} \right|}^2}}}.
\end{equation}
Note that the transmit power ${P_{r,1}}\left( {{t_1}} \right) \geq 0$ is positive in practical, thus ${{{{\left| {{h_{r,a}}} \right|}^2}{{\left| {{h_{a,r}}} \right|}^2} - 2^{\lambda _{fl}}2^{\lambda _{rl}}{{\left| {{\tilde{h}_{a,a}}} \right|}^2}{{\left| {{\tilde{h}_{r,r}}} \right|}^2}}}>0$. Given the channel data requirements, this imposes minimum requirement of the self-cancellation level.

Substitute (\ref{Pr1_FD_1TS}) into (\ref{Pb_FD_1TS}) and (\ref{Pu_FD_1TS}). We then have
\begin{equation}\label{Pb1_FD_1TS}
{P_{a,1}}\left( {{t_1}} \right) \approx \frac{{2^{\lambda _{fl}}\left( {2^{\lambda _{rl}}{{\left| {{\tilde{h}_{r,r}}} \right|}^2}\sigma _a^2 + {{\left| {{h_{r,a}}} \right|}^2}\sigma _r^2} \right)}}{{{{\left| {{h_{r,a}}} \right|}^2}{{\left| {{h_{a,r}}} \right|}^2} - 2^{\lambda _{fl}}2^{\lambda _{rl}}{{\left| {{\tilde{h}_{a,a}}} \right|}^2}{{\left| {{\tilde{h}_{r,r}}} \right|}^2}}},
\end{equation}
\begin{equation}\label{Pu1_FD_1TS}
{P_{b,1}}\left( {{t_1}} \right) \approx \frac{{2^{\lambda _{rl}}{{\left| {{h_{a,r}}} \right|}^2}\left( {2^{\lambda _{rl}}{{\left| {{\tilde{h}_{r,r}}} \right|}^2}\sigma _a^2 + {{\left| {{h_{r,a}}} \right|}^2}\sigma _r^2} \right)}}{{\left( {{{\left| {{h_{r,a}}} \right|}^2}{{\left| {{h_{a,r}}} \right|}^2} - 2^{\lambda _{fl}}2^{\lambda _{rl}}{{\left| {{\tilde{h}_{a,a}}} \right|}^2}{{\left| {{\tilde{h}_{r,r}}} \right|}^2}} \right){{\left| {{h_{b,r}}} \right|}^2}}}.
\end{equation}

\textit{CASE II}: In the case of ${P_{r,rl}} < {P_{r,fl}}$, (\ref{Rru-FD1TS}) meets the equality for the optimal solution. $P_r$ is solved through the equalities of (\ref{Rru-FD1TS}) and (\ref{Pu_FD_1TS}), as given by
\begin{equation}\label{Pr2_FD_1TS}
{P_{r,2}}\left( {{t_1}} \right) \hspace{-1mm} = \hspace{-1mm} P_{r,fl} \hspace{-1mm} \approx \hspace{-1mm} \frac{{2^{\lambda _{fl}}\left( {2^{\lambda _{rl}}{{\left| {{\tilde{h}_{b,b}}} \right|}^2}\sigma _r^2 \hspace{-1mm} + \hspace{-1mm} {{\left| {{h_{b,r}}} \right|}^2}\sigma _b^2} \right)}}{{{{\left| {{h_{r,b}}} \right|}^2}{{\left| {{h_{b,r}}} \right|}^2} \hspace{-1mm} - \hspace{-1mm} 2^{\lambda _{fl}}2^{\lambda _{rl}}{{\left| {{\tilde{h}_{b,b}}} \right|}^2}{{\left| {{\tilde{h}_{r,r}}} \right|}^2}}}.
\end{equation}
Substitute (\ref{Pr2_FD_1TS}) into (\ref{Pb_FD_1TS}) and (\ref{Pu_FD_1TS}). We then have
\begin{equation}\label{Pb2_FD_1TS}
{P_{a,2}}\left( {{t_1}} \right) \approx \frac{{2^{\lambda _{fl}}{{\left| {{h_{b,r}}} \right|}^2}\left( {2^{\lambda _{fl}}{{\left| {{\tilde{h}_{r,r}}} \right|}^2}\sigma _b^2 + {{\left| {{h_{r,b}}} \right|}^2}\sigma _r^2} \right)}}{{\left( {{{\left| {{h_{r,b}}} \right|}^2}{{\left| {{h_{b,r}}} \right|}^2} - 2^{\lambda _{fl}}2^{\lambda _{rl}}{{\left| {{\tilde{h}_{b,b}}} \right|}^2}{{\left| {{\tilde{h}_{r,r}}} \right|}^2}} \right){{\left| {{h_{a,r}}} \right|}^2}}},
\end{equation}
\begin{equation}\label{Pu2_FD_1TS}
{P_{b,2}}\left( {{t_1}} \right) \approx \frac{{2^{\lambda _{rl}}\left( {2^{\lambda _{fl}}{{\left| {{\tilde{h}_{r,r}}} \right|}^2}\sigma _b^2 + {{\left| {{h_{r,b}}} \right|}^2}\sigma _r^2} \right)}}{{{{\left| {{h_{r,b}}} \right|}^2}{{\left| {{h_{b,r}}} \right|}^2} - 2^{\lambda _{fl}}2^{\lambda _{rl}}{{\left| {{\tilde{h}_{b,b}}} \right|}^2}{{\left| {{\tilde{h}_{r,r}}} \right|}^2}}}.
\end{equation}

Now, we can obtain the equivalent optimization problem of (\ref{EE_Opt_FD_1TS}) in the same form as (\ref{EE_Opt_General}), with the objective changed to
\begin{equation}\label{Etotal_1TS}
E\left( {{t_1}} \right) = \max \left\{ {{E_{1}}\left( {{t_1}} \right),{E_{2}}\left( {{t_1}} \right)} \right\},
\end{equation}
where $E_1(t)$ and $E_2(t)$ are the energy consumption in \textit{CASE I} and \textit{CASE II}, respectively.

In the case of FD-TWR-1TS, $\mathbf{t}_{\min}=t_{\min,1}$. $t_{\min,1}$ can be obtained by setting \eqref{Pr1_FD_1TS}--\eqref{Pu2_FD_1TS} to be no larger than the corresponding maximum transmit powers, solving the inequalities, and choosing the intersection of the results. Specifically, we can rearrange the inequalities to be polynomial with all positive coefficients. Given the monotonicity of such polynomials, one-dimensional search, such as the bisection method, can be used to efficiently achieve $t_{\min,1}$.

\vspace{3mm}

\begin{theorem} \label{thtpa_FD_1TS}
\textit{Under TPA, (\ref{EE_Opt_General}) is convex for FD-TWR-1TS in the case of high data rate demands.}
\end{theorem}

\vspace{3mm}

\begin{IEEEproof}
In the asymptotic case that $2^{\lambda_{fl}}+2^{\lambda_{rl}}\gg 1$, exploiting the TPA model, ${E_{i}}\left( {{t_1}} \right)$ can be written as
\begin{align}
{E_{i}}\left( {{t_1}} \right) &= \sqrt {\frac{{{P_{\max ,a}}}}{{{\eta _{\max ,a}^2}}}} {t_1}\sqrt {{P_{a,i}}\left( {{t_1}} \right)} + \sqrt {\frac{{{P_{\max ,b}}}}{{{\eta _{\max ,b}^2}}}} {t_1}\sqrt {{P_{b,i}}\left( {{t_1}} \right)} \nonumber \\
&+ \sqrt {\frac{{{P_{\max ,r}}}}{{{\eta _{\max ,r}^2}}}} {t_1}\sqrt {{P_{r,i}}\left( {{t_1}} \right)} + P_3{t_1} + {P_{idle}}T, \nonumber
\end{align}
where $i\in \{1,2\}$ and $P_3 = {P_{rx}} + {P_{base}} + \varepsilon (R_{fl}+2 R_{rl}) - {P_{idle}}$.

${E_{i}}\left( {{t_1}} \right)$ is the positively weighted sum of ${t_1}\sqrt{P_{a,i}\left( {{t_1}} \right)}$, ${t_1}\sqrt{P_{b,i}\left( {{t_1}} \right)}$ and ${t_1}\sqrt{P_{r,i}\left( {{t_1}} \right)}$. We proceed to prove that each of these three components is convex.

\textit{CASE I}: In the case of $P_{r,rl}\geq P_{r,fl}$, ${t_1}\sqrt{P_{a,1}\left( {{t_1}} \right)}$ and ${t_1}\sqrt{P_{r,1}\left( {{t_1}} \right)}$ can be confirmed to be strictly convex, since
\begin{equation}\nonumber
\frac{{{\partial ^2}\left( {{t_1}\sqrt{P_{a,1}\left( {{t_1}} \right)}} \right)}}{{\partial {t_1}^2}} > 0\ , \ \frac{{{\partial ^2}\left( {{t_1}\sqrt{P_{r,1}\left( {{t_1}} \right)}} \right)}}{{\partial {t_1}^2}} > 0.
\end{equation}

Next, we prove the convexity of ${t_1}\sqrt{P_{b,1}\left( {{t_1}} \right)}$. To do this, we can rewrite ${t_1}\sqrt{P_{b,1}\left( {{t_1}} \right)}$ as
\begin{align}
&{t_1}\sqrt {{P_{b,1}}\left( {{t_1}} \right)} = \nonumber \\
&{t_1}\sqrt {\frac{{{2^{\frac{{{R_{rl}}T}}{{W{t_1}}}}}{{\left| {{h_{a,r}}} \right|}^2}\left( {{2^{\frac{{{R_{rl}}T}}{{W{t_1}}}}}{{\left| {{\tilde{h}_{r,r}}} \right|}^2}\sigma _a^2 \hspace{-1mm} + \hspace{-1mm} {{\left| {{h_{r,a}}} \right|}^2}\sigma _r^2} \right)}}{{\left( {{{\left| {{h_{r,a}}} \right|}^2}{{\left| {{h_{a,r}}} \right|}^2} \hspace{-1mm} - \hspace{-1mm} {2^{\frac{{\left( {{R_{rl}} + {R_{fl}}} \right)T}}{{W{t_1}}}}}{{\left| {{\tilde{h}_{a,a}}} \right|}^2}{{\left| {{\tilde{h}_{r,r}}} \right|}^2}} \right){{\left| {{h_{b,r}}} \right|}^2}}}} \nonumber \\
&= d_3 \times x\sqrt {\frac{{{2^{\frac{1}{x}}}\left( {1 + a_3{2^{\frac{1}{x}}}} \right)}}{{1 - b_3{2^{\frac{c_3}{x}}}}}}=d_3 \times f_{TPA}(x), \nonumber
\end{align}
where $x = (W{t_1})/({R_{rl}}T) > 0$, and for notational simplicity,
\begin{equation}
a_3=\frac{{{{\left| {{\tilde{h}_{r,r}}} \right|}^2}\sigma _a^2}}{{{{\left| {{h_{r,a}}} \right|}^2}\sigma _r^2}} > 0,\ b_3=\frac{{{{\left| {{\tilde{h}_{a,a}}} \right|}^2}{{\left| {{\tilde{h}_{r,r}}} \right|}^2}}}{{{{\left| {{h_{r,a}}} \right|}^2}{{\left| {{h_{a,r}}} \right|}^2}}} > 0, \nonumber
\end{equation}
\begin{equation}
c_3=\frac{{\left( {{R_{rl}} + {R_{fl}}} \right)}}{{{R_{rl}}}} > 1,\ d_3=\sqrt {\frac{{\sigma _r^2}}{{{{\left| {{h_{b,r}}} \right|}^2}}}} \frac{{{R_{rl}}T}}{W} > 0. \nonumber
\end{equation}

To prove the convexity of $t_1\sqrt{P_{b,1}\left( {{t_1}} \right)}$ now becomes to prove $f_{TPA}(x)$ is convex over $x\in(0,\infty)$. The second-order derivative of $f_{TPA}(x)$ can be given by
\begin{align}\label{DfTPA}
&\frac{{{\partial ^2}f_{TPA}\left( x \right)}}{{\partial {x^2}}} = {2^{ - 2 + \frac{1}{x}}}\log {{\left( 2 \right)}^2} \times \\
&\frac{{\left[ {{g_3}\left( c_3 \right) + {4^{\frac{1}{x}}}a_3{g_1}\left( c_3 \right) + {2^{1 + \frac{1}{x}}}a_3{g_2}\left( c_3 \right)} \right]}}{{\left( {1 + {2^{\frac{1}{x}}}a_3} \right)\sqrt {\frac{{{2^{\frac{1}{x}}}\left( {1 + a_3{2^{\frac{1}{x}}}} \right)}}{{1 - b_3{2^{\frac{c_3}{x}}}}}} {{\left( {1 - b_3{2^{\frac{c_3}{x}}}} \right)}^3}{x^3}}}, \nonumber
\end{align}
where the coefficients are obtained through mathematic manipulation, as given by
\begin{align}
&{g_1}\left( c_3 \right) \hspace{-0.5mm} = \hspace{-0.5mm} 4 \hspace{-0.5mm} + \hspace{-0.5mm} {4^{\frac{c_3}{x}}}{{b_3}^2}{\left( {c_3 \hspace{-0.5mm} - \hspace{-0.5mm} 2} \right)^2} \hspace{-0.5mm} + \hspace{-0.5mm} {2^{\frac{{c_3 + x}}{x}}}b_3\left( {{{c_3}^2} \hspace{-0.5mm} + \hspace{-0.5mm} 2c_3 \hspace{-0.5mm} - \hspace{-0.5mm} 4} \right), \nonumber \\
&{g_2}\left( c_3 \right) \hspace{-0.5mm} = \hspace{-0.5mm} 3 \hspace{-0.5mm} + \hspace{-0.5mm} {4^{\frac{c_3}{x}}}{{b_3}^2}\left( {{{c_3}^2} \hspace{-0.5mm} - \hspace{-0.5mm} 3c_3 \hspace{-0.5mm} + \hspace{-0.5mm} 3} \right) \hspace{-0.5mm} + \hspace{-0.5mm} {2^{\frac{c_3}{x}}}b_3\left( {2{{c_3}^2} \hspace{-0.5mm} + \hspace{-0.5mm} 3c_3 \hspace{-0.5mm} - \hspace{-0.5mm} 6} \right), \nonumber \\
&{g_3}\left( c_3 \right) \hspace{-0.5mm} = \hspace{-0.5mm} {1 \hspace{-0.5mm} + \hspace{-0.5mm} {4^{\frac{c_3}{x}}}{{b_3}^2}{{\left( {c_3 \hspace{-0.5mm} - \hspace{-0.5mm} 1} \right)}^2} \hspace{-0.5mm} + \hspace{-0.5mm} {2^{\frac{{c_3 + x}}{x}}}b_3\left( {{{c_3}^2} \hspace{-0.5mm} + \hspace{-0.5mm} c_3 \hspace{-0.5mm} - \hspace{-0.5mm} 1} \right)}. \nonumber
\end{align}

We first set $c_3^2+2c_3-4>0$, $c_3^2-3c_3+3>0$, $2c_3^2+3c_3-6>0$, and $c_3^2+c_3-1>0$, and then solve the inequalities. We see that all the four polynomials are positive for $c_3 \in (\sqrt{5}-1,\infty)$. We prove that the second-order derivative of $f_{TPA}(x)$ is positive over $c_3 \in (\sqrt{5}-1,\infty)$.

We can also meticulously rewrite $g_1(c_3)$ and $g_2(c_3)$ as
\begin{align}
&{g_1}\left( c_3 \right) = {\left( {c_3 - 2} \right)^2} \times \nonumber \\
&\left( {{{\left( {{2^{\frac{c_3}{x}}}b_3 \hspace{-1mm} + \hspace{-1mm} \frac{{{{c_3}^2} + 4c_3 - 8}}{{2{{\left( {c_3 - 2} \right)}^2}}}} \right)}^2} \hspace{-1mm} - \hspace{-1mm} \frac{{{{c_3}^2}\left( {{{c_3}^2} + 4c_3 - 8} \right)}}{{{{\left( {c_3 - 2} \right)}^4}}}} \right), \nonumber
\end{align}
\begin{align}
&{g_2}\left( c_3 \right) = \left( {{{c_3}^2} - 3c_3 + 3} \right) \times \nonumber \\
&\left( {{{\left( {{2^{\frac{c_3}{x}}}b_3 + \frac{{2{{c_3}^2} \hspace{-1mm} + \hspace{-1mm} 3c_3 - 6}}{{2\left( {{{c_3}^2} - 3c_3 + 3} \right)}}} \right)}^2} \hspace{-1mm} - \hspace{-1mm} \frac{{{{c_3}^2}\left( {4{{c_3}^2} \hspace{-1mm} + \hspace{-1mm} 12c_3 \hspace{-1mm} - \hspace{-1mm} 27} \right)}}{{4{{\left( {{{c_3}^2} - 3c_3 + 3} \right)}^2}}}} \right). \nonumber
\end{align}

A sufficient condition for $g_i(c_3),\ i \in \{1,2,3\}$, all being positive, is that ${{c_3}^2} + 4c_3 - 8 < 0$, $4{c_3}^2 + 12c_3 - 27 < 0$, and ${c_3}^2+c_3-1 > 0$. We can solve the inequalities and obtain the feasible region, $c_3 \in (1,2\sqrt{3}-2)$, given $c_3>1$. In other words, the second-order derivative of $f_{TPA}(x)$ is positive when $c_3 \in (1,2\sqrt{3}-2)$.

As a result, we prove $\frac{{{\partial ^2}f_{TPA}\left( x \right)}}{{\partial {x^2}}} > 0$ over $c_3 \in (1,2\sqrt{3}-2)\bigcup(\sqrt{5}-1,\infty)=(1,\infty)$. Therefore, $f_{TPA}(x)$ is convex. In turn, ${t_1}\sqrt{P_{b,1}\left( {{t_1}} \right)}$ is convex, and so is $E_{1}(t_1)$, the positively weighted sum of ${t_1}\sqrt{P_{a,1}\left( {{t_1}} \right)}$, ${t_1}\sqrt{P_{b,1}\left( {{t_1}} \right)}$ and ${t_1}\sqrt{P_{r,1}\left( {{t_1}} \right)}$.

\textit{CASE II}: In the case of $P_{r,rl} < P_{r,fl}$, ${t_1}\sqrt{{P_{b,2}}}$ and ${t_1}\sqrt{{P_{r,2}}}$ are convex, as proved in \textit{CASE I}. We rewrite ${t_1}\sqrt{{P_{a,2}}}$ as
\begin{align}
&{t_1}\sqrt {{P_{a,2}}\left( {{t_1}} \right)}  = \nonumber \\
&{t_1}\sqrt {\frac{{{2^{\frac{{{R_{fl}}T}}{{W{t_1}}}}}{{\left| {{h_{b,r}}} \right|}^2}\left( {{2^{\frac{{{R_{fl}}T}}{{W{t_1}}}}}{{\left| {{\tilde{h}_{r,r}}} \right|}^2}\sigma _b^2 \hspace{-1mm} + \hspace{-1mm} {{\left| {{h_{r,b}}} \right|}^2}\sigma _r^2} \right)}}{{\left( {{{\left| {{h_{r,b}}} \right|}^2}{{\left| {{h_{b,r}}} \right|}^2} \hspace{-1mm} - \hspace{-1mm} {2^{\frac{{\left( {{R_{rl}} + {R_{fl}}} \right)T}}{{W{t_1}}}}}{{\left| {{\tilde{h}_{b,b}}} \right|}^2}{{\left| {{\tilde{h}_{r,r}}} \right|}^2}} \right){{\left| {{h_{a,r}}} \right|}^2}}}} \nonumber \\
&= \sqrt {\frac{{\sigma _r^2}}{{{{\left| {{h_{a,r}}} \right|}^2}}}} \frac{{{R_{fl}}T}}{W}y\sqrt {\frac{{{2^{\frac{1}{y}}}\left( {1 + \frac{{{{\left| {{\tilde{h}_{r,r}}} \right|}^2}\sigma _b^2}}{{{{\left| {{h_{r,b}}} \right|}^2}\sigma _r^2}}{2^{\frac{1}{y}}}} \right)}}{{1 - {2^{\frac{{\left( {{R_{rl}} + {R_{fl}}} \right)}}{{{R_{fl}}y}}}}\frac{{{{\left| {{\tilde{h}_{b,b}}} \right|}^2}{{\left| {{\tilde{h}_{r,r}}} \right|}^2}}}{{{{\left| {{h_{r,b}}} \right|}^2}{{\left| {{h_{b,r}}} \right|}^2}}}}}}, \nonumber
\end{align}
where $y = (W{t_1})/({R_{fl}}T) > 0$ for notational simplicity.

We can see that $t_1\sqrt{P_{a,2}(t_1})$ has the same structure as $t_1\sqrt{P_{b,1}(t_1})$. The only difference is the positive coefficients. To this end, we can evaluate the convexity of $t_1\sqrt{P_{a,2}(t_1})$ in the same way, as we did on $t_1\sqrt{P_{b,1}(t_1})$, and prove the convexity of $t_1\sqrt{P_{a,2}(t_1})$. The detailed proof is suppressed due to limited space.

Given that  both ${E_{1}}\left( {{t_1}} \right)$ and ${E_{2}}\left( {{t_1}} \right)$ are convex, $E\left( {{t_1}} \right) = \max \left\{ {{E_{1}}\left( {{t_1}} \right),{E_{2}}\left( {{t_1}} \right)} \right\}$ is convex under the TPA model.
\end{IEEEproof}

\vspace{3mm}

\begin{theorem} \label{thetpa_FD_1TS}
\textit{Under ETPA, (\ref{EE_Opt_General}) is convex for FD-TWR-1TS in the case of high data rate demands.}
\end{theorem}

\vspace{3mm}

\begin{IEEEproof}
In the asymptotic case that $2^{\lambda_{fl}}+2^{\lambda_{rl}}\gg 1$, exploiting the ETPA model, $E_i(t_1)$ can be written as
\begin{align}
{E_{i}}\left( {{t_1}} \right) &= \frac{{{t_1}{P_{a,i}}\left( {{t_1}} \right)}}{{\left( {1 + u \kappa_a} \right){\eta _{\max ,a}}}} + \frac{{{t_1}{P_{b,i}}\left( {{t_1}} \right)}}{{\left( {1 + u \kappa_b} \right){\eta _{\max ,b}}}}  \nonumber \\
&+\frac{{{t_1}{P_{r,i}}\left( {{t_1}} \right)}}{{\left( {1 + u \kappa_r} \right){\eta _{\max ,r}}}} + P_4{t_1} + {P_{idle}}T, \nonumber
\end{align}
where $i \in \{1,2\}$ and $P_4 = \frac{{u \kappa_a {P_{\max ,a}}}}{{\left( {1 + u \kappa_a} \right){\eta _{\max ,a}}}} + \frac{{u \kappa_b {P_{\max ,b}}}}{{\left( {1 + u \kappa_b} \right){\eta _{\max ,b}}}} + \frac{{u \kappa_r {P_{\max ,r}}}}{{\left( {1 + u \kappa_r} \right){\eta _{\max ,r}}}} + {P_{rx}} + P_{base} + \varepsilon \left(R_{fl}+2 R_{rl}\right) - {P_{idle}}$.

${E_{i}}\left( {{t_1}} \right)$ is the positively weighted sum of ${t_1}{P_{a,i}\left( {{t_1}} \right)}$, ${t_1}{P_{b,i}\left( {{t_1}} \right)}$ and ${t_1}{P_{r,i}\left( {{t_1}} \right)}$. We proceed to prove that each of these three components is convex.

\textit{CASE I}: In the case of $P_{r,rl}\geq P_{r,fl}$, ${t_1}{P_{a,1}\left( {{t_1}} \right)}$ and ${t_1}{P_{r,1}\left( {{t_1}} \right)}$ are both convex, as can be readily extended from the proof of Theorem~\ref{thtpa_FD_1TS}. ${t_1}{P_{b,1}\left( {{t_1}} \right)}$ can be written as
\begin{align}
&{t_1}{P_{b,1}}\left( {{t_1}} \right) = \nonumber \\
&{t_1}\frac{{{2^{\frac{{{R_{rl}}T}}{{W{t_1}}}}}{{\left| {{h_{a,r}}} \right|}^2}\left( {{2^{\frac{{{R_{rl}}T}}{{W{t_1}}}}}{{\left| {{\tilde{h}_{r,r}}} \right|}^2}\sigma _a^2 + {{\left| {{h_{r,a}}} \right|}^2}\sigma _r^2} \right)}}{{\left( {{{\left| {{h_{r,a}}} \right|}^2}{{\left| {{h_{a,r}}} \right|}^2} \hspace{-1mm} - \hspace{-1mm} {2^{\frac{{\left( {{R_{rl}} + {R_{fl}}} \right)T}}{{W{t_1}}}}}{{\left| {{\tilde{h}_{a,a}}} \right|}^2}{{\left| {{\tilde{h}_{r,r}}} \right|}^2}} \right){{\left| {{h_{b,r}}} \right|}^2}}} \nonumber \\
&= e_4 \times x{\frac{{{2^{\frac{1}{x}}}\left( {1 + a_4{2^{\frac{1}{x}}}} \right)}}{{1 - b_4{2^{\frac{c_4}{x}}}}}}=e_4 \times f_{ETPA}(x), \nonumber
\end{align}
where $x = (W{t_1})/({R_{rl}}T) > 0$, and for notational simplicity,
\begin{equation}\nonumber
a_4=\frac{{{{\left| {{\tilde{h}_{r,r}}} \right|}^2}\sigma _a^2}}{{{{\left| {{h_{r,a}}} \right|}^2}\sigma _r^2}} > 0,\ b_4=\frac{{{{\left| {{\tilde{h}_{a,a}}} \right|}^2}{{\left| {{\tilde{h}_{r,r}}} \right|}^2}}}{{{{\left| {{h_{r,a}}} \right|}^2}{{\left| {{h_{a,r}}} \right|}^2}}} > 0,
\end{equation}
\begin{equation}\nonumber
c_4=\frac{{\left( {{R_{rl}} + {R_{fl}}} \right)}}{{{R_{rl}}}} > 1,\ e_4={\frac{{\sigma _r^2}}{{{{\left| {{h_{b,r}}} \right|}^2}}}} \frac{{{R_{rl}}T}}{W} > 0.
\end{equation}

To prove the convexity of ${t_1}{P_{b,1}}\left( {{t_1}} \right)$ now becomes to prove $f_{ETPA}(x)$ is convex over $x\in(0,\infty)$. The second-order derivative of $f_{ETPA}(x)$ can be expressed as
\begin{align}\label{DfETPA}
&\frac{{{\partial ^2}f_{ETPA}\left( x \right)}}{{\partial {x^2}}} = {2^{\frac{1}{x}}}\log {{\left( 2 \right)}^2} \times \nonumber \\
&\frac{{\left[ {1 \hspace{-1mm} + \hspace{-1mm} {4^{\frac{c_4}{x}}}{{b_4}^2}{{\left( {c_4 \hspace{-1mm} - \hspace{-1mm} 1} \right)}^2} \hspace{-1mm} + \hspace{-1mm} {2^{\frac{c_4}{x}}}b_4\left( {{{c_4}^2} \hspace{-1mm} + \hspace{-1mm} 2c_4 \hspace{-1mm} - \hspace{-1mm} 2} \right) \hspace{-1mm} + \hspace{-1mm} {2^{\frac{1}{x}}}a_4{g_4}\left( c_4 \right)} \right]}}{{{{\left( {1 - b_4{2^{\frac{c_4}{x}}}} \right)}^3}{x^3}}},
\end{align}
where ${g_4}\left( c_4 \right)$ is given by
\begin{equation}\nonumber
{g_4}\left( c_4 \right) = 4 + {4^{\frac{c_4}{x}}}{{b_4}^2}{\left( {c_4 - 2} \right)^2} + {2^{\frac{c_4}{x}}}b_4\left( {{{c_4}^2} + 4c_4 - 8} \right).
\end{equation}

We first set ${{c_4}^2} + 2c_4 - 2>0$ and ${{c_4}^2} + 4c_4 - 8>0$, and solve the inequalities. We obtain $c_4 \in (2\sqrt{3}-2,\infty)$; in other words, we prove that the second-order derivative of $f_{ETPA}(x)$ is positive over $c_4 \in (2\sqrt{3}-2,\infty)$.

We can also rewrite $g_4(c_4)$ as
\begin{align}
&{g_4}\left( c_4 \right) = {\left( {c_4 - 2} \right)^2} \times \nonumber \\
&\left( {{{\left( {{2^{\frac{c_4}{x}}}b_4 + \frac{{{{c_4}^2} \hspace{-1mm} + \hspace{-1mm} 4c_4 - 8}}{{2{{\left( {c_4 - 2} \right)}^2}}}} \right)}^2} \hspace{-1mm} - \hspace{-1mm} \frac{{{{c_4}^2}\left( {{{c_4}^2} + 8c_4 - 16} \right)}}{{4{{\left( {c_4 - 2} \right)}^4}}}} \right). \nonumber
\end{align}

We then set ${{c_4}^2} + 2c_4 - 2>0$ and ${{c_4}^2} + 8c_4 - 16<0$. We solve the inequalities and obtain $c_4 \in (1,4\sqrt{2}-4)$, given $c_4>1$. In other words, the second-order derivative of $f_{ETPA}(x)$ is positive for $c_4 \in (1,4\sqrt{2}-4)$.

As a result, $\frac{{{\partial ^2}f_{ETPA}\left( x \right)}}{{\partial {x^2}}} > 0$ over $c_4 \in (1,4\sqrt{2}-4)\bigcup (2\sqrt{3}-2,\infty) = (1,\infty)$. $f_{ETPA}(x)$ is convex. Therefore, ${t_1}{P_{b,1}}\left( {{t_1}} \right)$ is convex, and so is $E_{1}(t_1)$, as well.

\textit{CASE II}: In the case of $P_{r,rl}<P_{r,fl}$, ${t_1}{P_{b,2}\left( {{t_1}} \right)}$ and ${t_1}{P_{r,2}\left( {{t_1}} \right)}$ are convex due to the fact that the second-order derivatives of these are positive. ${t_1}{P_{a,2}\left( {{t_1}} \right)}$ can be rewritten as given by
\begin{align}
&{t_1}{P_{a,2}}\left( {{t_1}} \right) = \nonumber \\
&{t_1}\frac{{{2^{\frac{{{R_{fl}}T}}{{W{t_1}}}}}{{\left| {{h_{b,r}}} \right|}^2}\left( {{2^{\frac{{{R_{fl}}T}}{{W{t_1}}}}}{{\left| {{\tilde{h}_{r,r}}} \right|}^2}\sigma _b^2 + {{\left| {{h_{r,b}}} \right|}^2}\sigma _r^2} \right)}}{{\left( {{{\left| {{h_{r,b}}} \right|}^2}{{\left| {{h_{b,r}}} \right|}^2} \hspace{-1mm} - \hspace{-1mm} {2^{\frac{{\left( {{R_{rl}} + {R_{fl}}} \right)T}}{{W{t_1}}}}}{{\left| {{\tilde{h}_{b,b}}} \right|}^2}{{\left| {{\tilde{h}_{r,r}}} \right|}^2}} \right){{\left| {{h_{a,r}}} \right|}^2}}} \nonumber \\
&= \frac{{\sigma _r^2}}{{{{\left| {{h_{a,r}}} \right|}^2}}}\frac{{{R_{fl}}T}}{W}y\frac{{{2^{\frac{1}{y}}}\left( {1 + \frac{{{{\left| {{\tilde{h}_{r,r}}} \right|}^2}\sigma _b^2}}{{{{\left| {{h_{r,b}}} \right|}^2}\sigma _r^2}}{2^{\frac{1}{y}}}} \right)}}{{1 - {2^{\frac{{\left( {{R_{rl}} + {R_{fl}}} \right)}}{{{R_{fl}}y}}}}\frac{{{{\left| {{\tilde{h}_{b,b}}} \right|}^2}{{\left| {{\tilde{h}_{r,r}}} \right|}^2}}}{{{{\left| {{h_{r,b}}} \right|}^2}{{\left| {{h_{b,r}}} \right|}^2}}}}}, \nonumber
\end{align}
where $y = \frac{{W{t_1}}}{{{R_{fl}}T}} > 0$.

We see that $t_1{P_{a,2}(t_1})$ has the same structure as $t_1{P_{b,1}(t_1})$. The only difference is the positive coefficients. We can evaluate the convexity of $t_1{P_{a,2}(t_1})$ in the same way, as we have done on $t_1{P_{b,1}(t_1})$, and prove the convexity of $t_1{P_{a,2}(t_1})$. The detailed proof is suppressed due to limited space.

Given that both ${E_{1}}\left( {{t_1}} \right)$ and ${E_{2}}\left( {{t_1}} \right)$ are convex, $E\left( {{t_1}} \right) = \max \left\{ {{E_{1}}\left( {{t_1}} \right),{E_{2}}\left( {{t_1}} \right)} \right\}$ is convex under the ETPA model.
\end{IEEEproof}
\section{Simulation Results}\label{Sec6}
In this section, simulations are carried out to evaluate the maximum EE of FD-TWR. As dictated in Theorems \ref{thtpa_FD_2TS}, \ref{thetpa_FD_2TS}, \ref{thtpa_FD_1TS} and \ref{thetpa_FD_1TS}, the non-convex EE maximization of FD-TWR under non-ideal PAs and non-negligible circuit power can be reformulated to convex optimizations with optimality preserved and variables reduced ({i.e.}, w.r.t. the transmit durations only). The convex equivalents can be solved using standard solvers, such as the barrier method, the exterior penalty method and the sequential quadratic programming~(SQP) method. For illustration purpose, we use the barrier method.

For comparison purpose, we also simulate the maximum EE of HD-TWR (the details are provided in Appendix \ref{Apx3}). The frame duration is $T=10$ ms, unless otherwise specified. The PA efficiency $\eta_{\max,i}$ is set to $0.35$~\cite{PC_model}, which is within the reasonable range between 0.311 and 0.388~\cite{Reference40}. Reciprocal channels are assumed between every pair of nodes, {i.e.}, $h_{i,j}=h_{j,i}$ for all $i \neq j$. The noise is independent but identically distributed (i.i.d.) at all the nodes, {i.e.}, $\sigma_i^2=\sigma^2 ,\ \forall i$. The PA consumption simulated is in the order of dozens of Watts~\cite{PC_model}, depending on the distance of the transmitter and receiver, as well as the efficiency of the PA. The circuit power ranges from tens to hundreds of milliWatts, comprising mainly baseband processing and RF generation~\cite{PC_model}. Other simulation parameters are specified in Table~I, with reference to other works~\cite{Reference43,Reference24,Reference11} and 3GPP LTE specifications~\cite{3GPP}. For fair comparison, we ensure that all schemes involved achieve the same required data rates in both link directions over the fixed time frame $T$.

\begin{table}[htbp]\label{Table_Simulation}
\centering
\caption{Simulation Parameters}
\vspace{-2mm}
\begin{tabular}{|l|l|}
\hline
\textbf{Parameters} & \textbf{Values} \\
\hline
System bandwidth $W$ & 10\,MHz \\
Noise power spectral density ($N_0$) & --\,174\,dBm/Hz \\
Distance between nodes ($d_{a,r}$, $d_{r,b}$) & 50, 50\,m \\
Distance between antennas ($d_{a,a}$, $d_{b,b}$, $d_{r,r}$) & 5, 5, 5\,cm \\
Average path loss between nodes in dB ($\text{L}_{i,j}$) & $103.8+21\log_{10}d$ \\
Cancellation amount ($\alpha_{PS}$, $\alpha_{PSAC}$) & 40, 60\,dB \\
Idle power consumption ($P_{\text{idle},a}$,$P_{\text{idle},r}$,$P_{\text{idle},b}$) & 30, 15, 5\,mW \\
Static circuit power consumption ($P_{base,i}$) & 100, 50, 20\,mW \\
Dynamic circuit factor ($\varepsilon$) & 50\,mW/Gbps \\
Maximum output power ($P_{\max,i}$) & 46, 37, 23\,dBm \\
Maximum PA efficiency ($\eta_{\max,i}$) & 0.35, 0.35, 0.35 \\
\hline
\end{tabular}
\end{table}

\begin{figure}[!t]
\centering
\hspace{-4mm}
\includegraphics[scale=0.28]{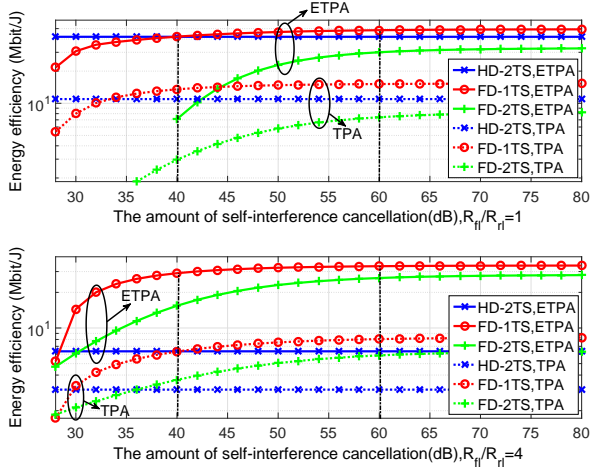}
\vspace{-3mm}
\caption{The optimal EE versus the self-cancellation capability of FD-TWR, where $R_{fl}+R_{rl}=65$ Mbps.}
\label{fig_SIC_con}
\end{figure}
Fig.~\ref{fig_SIC_con} plots the maximum EE of FD-TWR, as the self-cancellation capability of FD-TWR improves. The HD-TWR is plotted in blue for reference purpose. We can see FD-TWR-1TS can substantially outperform FD-TWR-2TS under a wide spectrum of self-cancellation capability, while the gap between the two schemes decreases as the self-cancellation improves. As shown in the figures, when the self-cancellation level is high, {i.e.}, the self-interference becomes negligible, the EEs of both FD-TWR-1TS and -2TS stabilize and flat out. As the self-cancellation level decreases, FD-TWR-2TS exhibits faster degradation than FD-TWR-1TS. This is because FD-TWR-1TS allows every node to transmit over the entire time frame. The transmit power of the nodes can be lower than that in FD-TWR-2TS, and in turn, produces less self-interference. In this sense, weak self-cancellation capability can still keep the nodes to receive properly in FD-TWR-1TS. In contrast, the transmit durations of the source/destination nodes are much shorter in FD-TWR-2TS. The required transmit powers of the nodes are much higher, due to the exponential-linear tradeoff between the transmit power and time. This results in strong self-interference, which cannot be properly cancelled at the low self-cancellation level. In other words, FD-TWR-2TS is more sensitive to the self-cancellation level.

We notice that HD-TWR can be more energy efficient than FD-TWR, when the relay and the source/destination nodes have poor self-cancellation capability, {e.g.}, $\leq 40$ dB if the traffic is balanced in the forward and reverse links, as shown in the upper figure of Fig.~\ref{fig_SIC_con}. The reason is that the residual self-interference is too large and results in a substantial loss of data rate. In turn, the EE of FD-TWR degrades.

We also notice that unbalanced traffic deteriorates the EE degradation, and HD-TWR is much more susceptible to the traffic imbalance than FD-TWR. For FD-TWR-1TS, this is because the transmission and reception of all the FD nodes are aligned within the same time duration. The EE of the unbalanced links decreases, because of excessively high transmit power over relatively short time in the high-rate direction, and excessively long transmission time incurring excessive circuit energy consumption in the low rate direction. For HD-TWR, the transmissions of the relay to both source/destination nodes also need to be aligned. Compared to FD-TWR-1TS, HD-TWR has much shorter time durations in which transmissions are aligned, incurring higher transmit power and lower EE, as shown in the lower figure of Fig.~\ref{fig_SIC_con}. Interestingly, FD-TWR-2TS is not much affected by the traffic imbalance. This is because the FD relay receives and relays the same traffic from one node at any instant. The increasingly unbalanced traffic only enlarges difference of time durations allocated to the two link directions at the relay, and does not affect the transmit power of the relay.

\begin{figure}[!t]
\centering
\hspace{-4mm}
\includegraphics[scale=0.28]{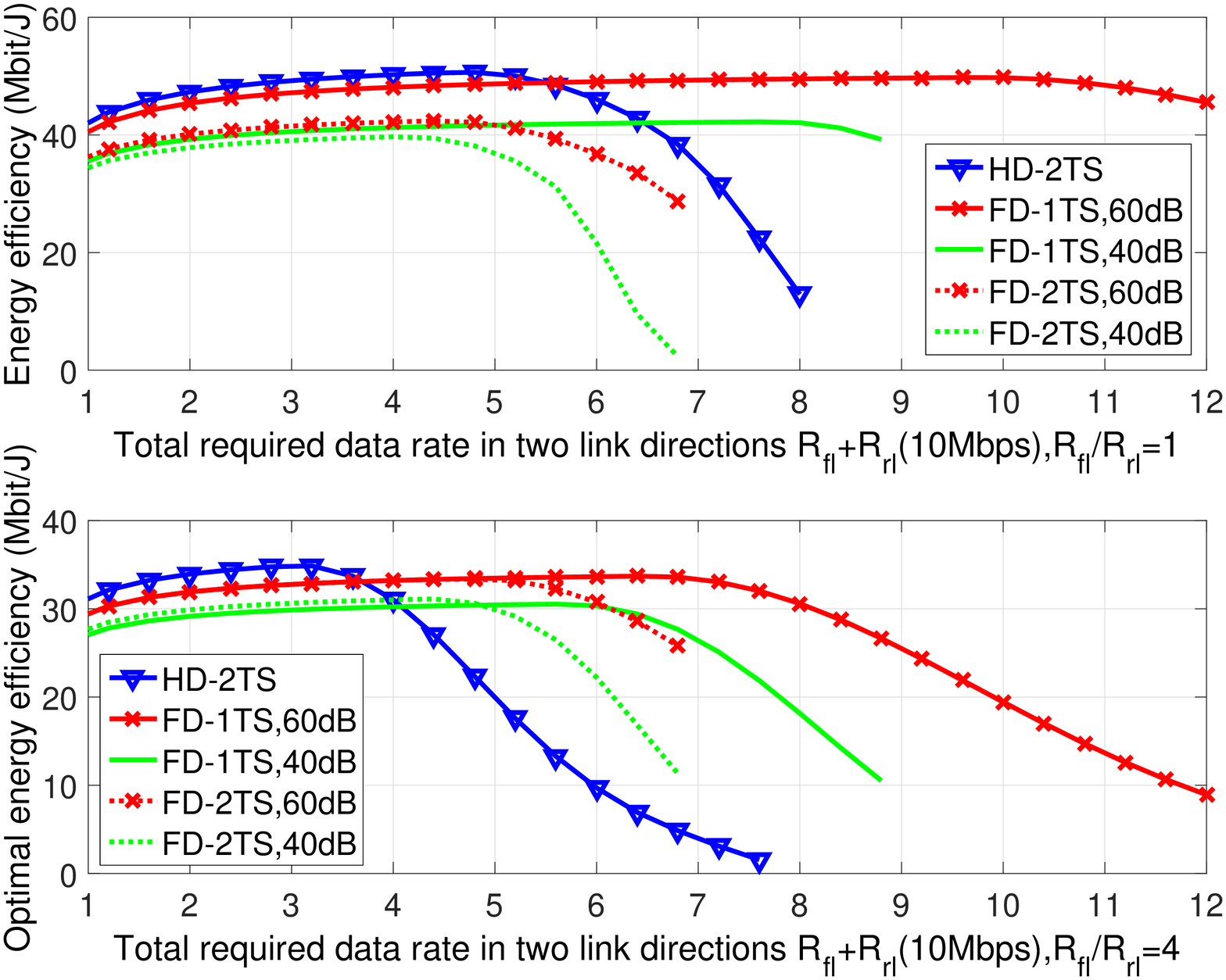}
\vspace{-3mm}
\caption{EE comparison between different transmission strategies w.r.t data rate demand under the ETPA model.}
\label{fig_SIC_R}
\end{figure}
Fig.~\ref{fig_SIC_R} plots the maximum EE of FD-TWR-1TS and -2TS, as well as HD-TWR, as the total rate requirement increases under the self-cancellation levels of 40 and 60 dB. We see that FD-TWR is superior to HD-TWR in terms of EE when the data requirement is high. Particularly, FD-TWR-1TS can maintain the same EE of 50 Mbits/J as HD-TWR, while doubling the data rate region of HD-TWR, from 55 Mbps to 110 Mbps. We also see that the self-cancellation level plays an important role in FD-TWR. Not only does it affect the maximum EE, but also affects the data rate that can be supported by the network. Reducing the level leads to a loss of both the EE and data rate. The traffic imbalance between the two link directions can further reduce the EE and data rate.

An interesting finding in Fig.~\ref{fig_SIC_R} is that HD-TWR can outperform both the FD-TWR schemes, when the data requirement is low. This is the case where the traffic demand is low and the radio resource is abundant. Only part of the time frame is allocated to data transmission in HD-TWR, and the energy can be most efficiently planned to leverage the PA consumption, circuit consumption, and data transmission. In contrast, FD-TWR has to align the transmission and reception of the FD nodes within the same time duration, incurring the loss of EE, as discussed in Fig.~\ref{fig_SIC_con}.

\begin{figure}[!t]
\centering
\subfloat[The optimized transmit time under the ETPA model]{\hspace{-4mm} \label{fig_non_impact_t} \includegraphics[scale=0.28]{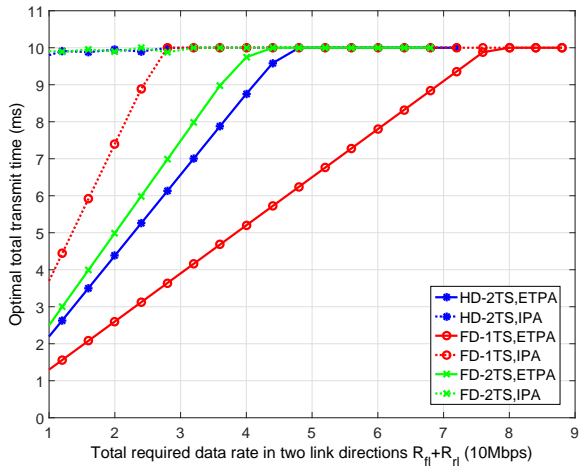}}
\vspace{-3mm}
\hfil
\centering
\subfloat[The optimized transmit power under the ETPA model]{\hspace{-4mm} \label{fig_non_impact_P} \includegraphics[scale=0.28]{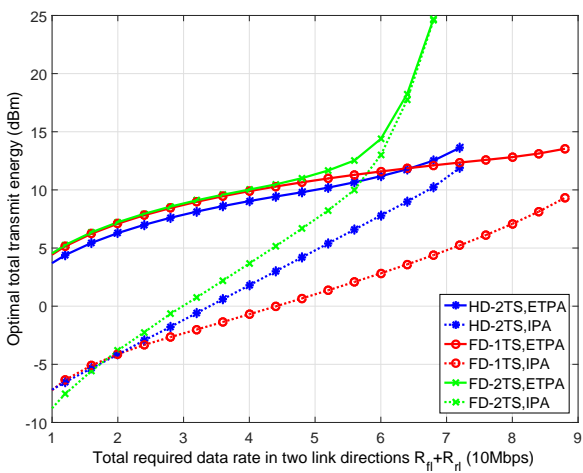}}
\caption{The optimal resource allocation results under non-ideal PA and non-negligible circuit power, where PS ($\alpha_{PS}=40$ dB) is adopted and $R_{fl}/R_{rl}=1$.}
\label{fig_non_impact}
\end{figure}
Fig.~\ref{fig_non_impact} corroborates the above findings by evaluating the optimized transmit powers and time durations. We see in Fig.~\ref{fig_non_impact}(a) that, when the traffic demand is less than 50 Mbps, the average transmit time within a frame is less than the entire frame for HD-TWR. This is the case where HD-TWR has the chance to separately optimize the transmit power of each link (the transmit power is yet to increase exponentially, as evident from Fig.~\ref{fig_non_impact}(b)), achieving the higher EE than FD-TWR.

We note that the optimal transmit time under non-negligible transmission-dependent circuit power is less than the entire time frame under low data requirements. The conclusion drawn is that the proposed algorithm can optimize the transmit time as such that the energy consumed for data transmission (including the PA consumption and transmission-dependent circuit operation) is leveraged with the energy consumed in the idle-state circuit.

We also notice that the optimal transmit time of FD-TWR-1TS is not always across the entire frame, even in the ideal case of negligible circuit power and ideal PAs. The reason is that FD-TWR-1TS needs to operate at a relatively high transmit power to overcome both the self-interference and noise. If the transmission lasts over the entire frame, the transmit power would be lower. The self-interference would also decrease. The receiver noise would become dominant over the self-interference, which could require extra transmit power to suppress the noise given the required data rate and thus reduce the EE.

\begin{figure*}[!t]
\centering
\subfloat[The impact of PA efficiency and traffic imbalance where PS ($\alpha_{PS}=40$ dB) is adopted.]{\hspace{-4mm} \label{Ratio_PA_3D} \includegraphics[scale=0.19]{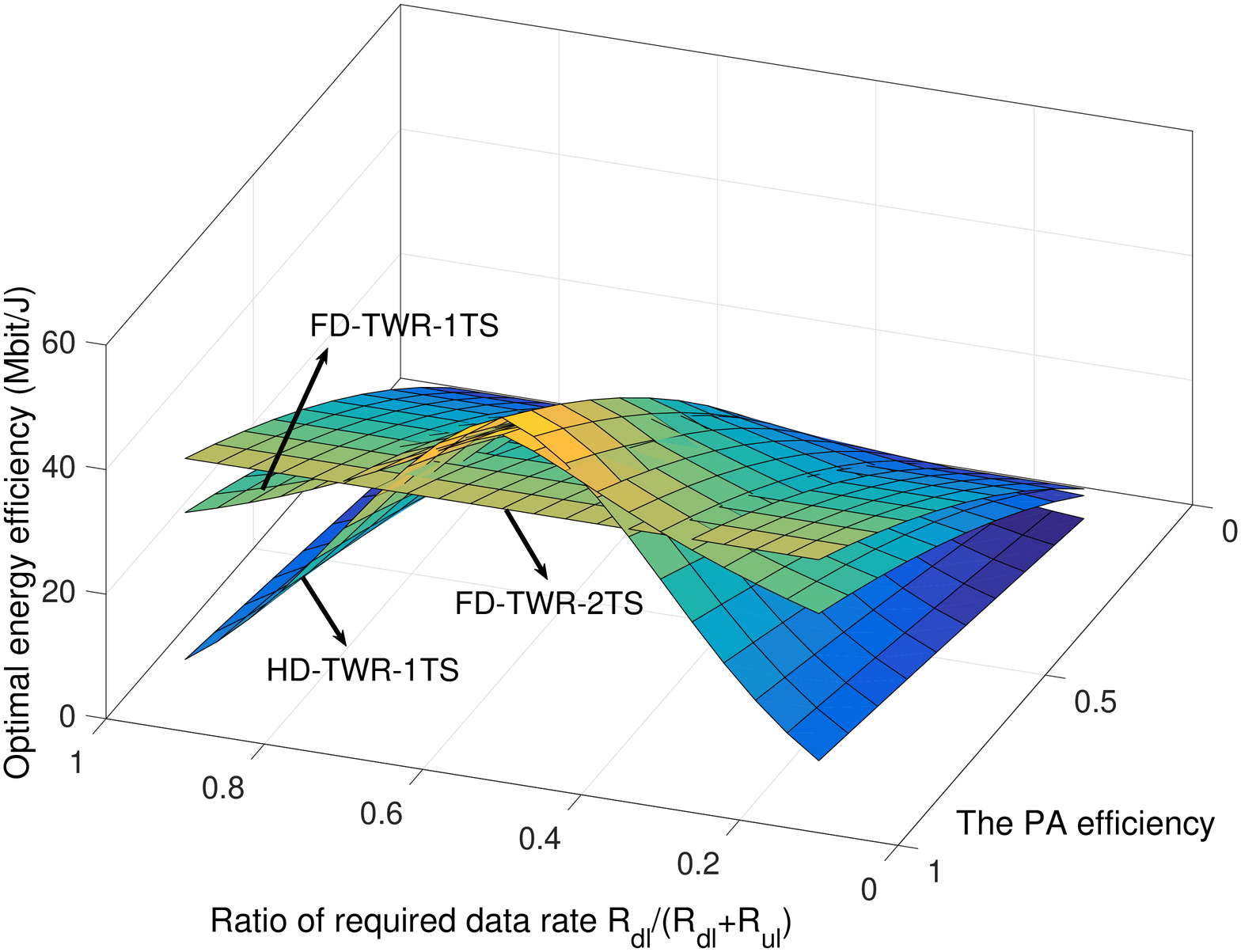}}
\hfil
\centering
\subfloat[The impact of traffic imbalance and self-cancellation.]{\hspace{-4mm} \label{SI_Ratio_3D} \includegraphics[scale=0.19]{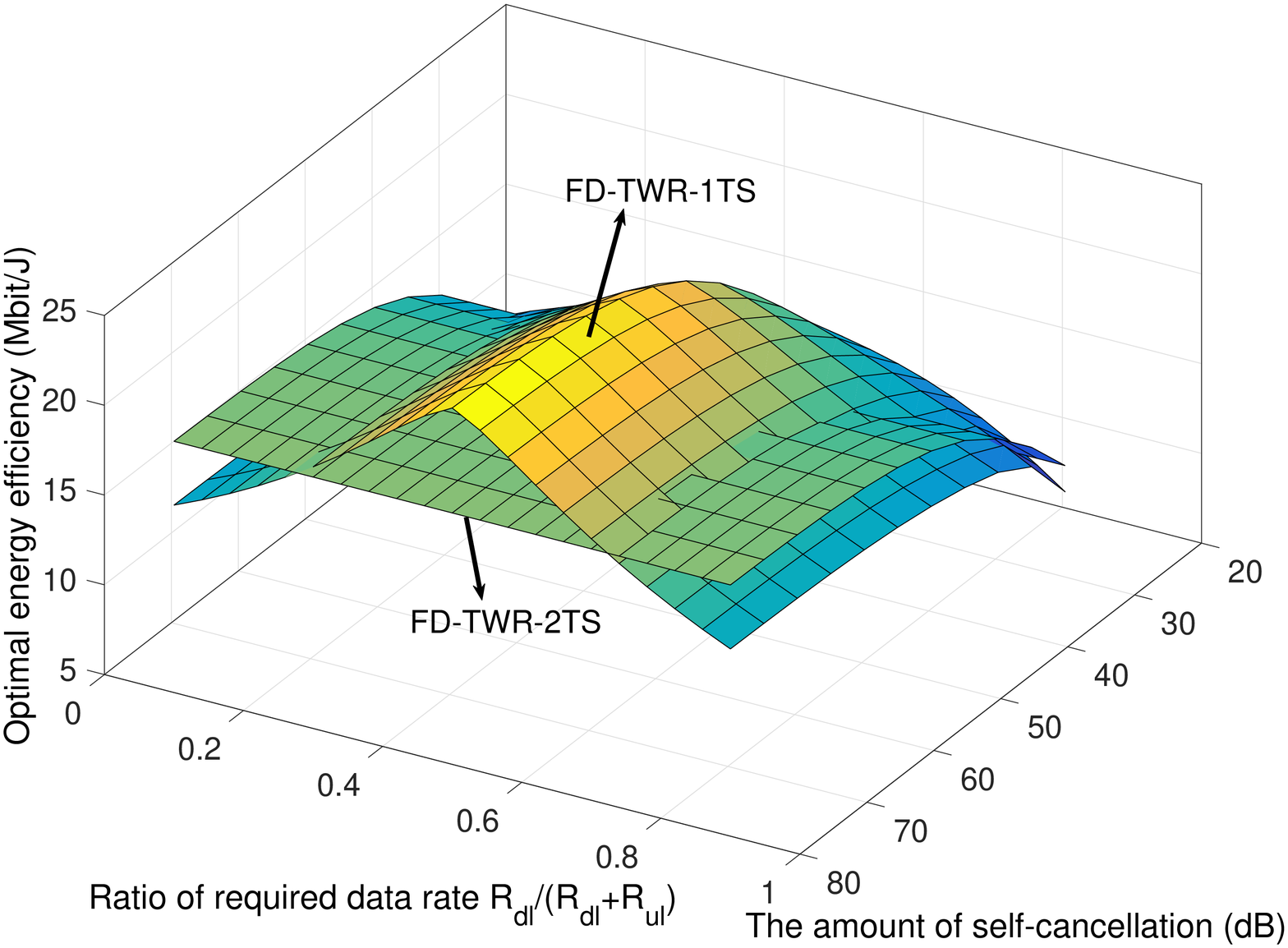}}
\hfil
\centering
\subfloat[The impact of PA efficiency and self-cancellation under symmetric traffic.]{\hspace{-4mm} \label{SI_PA_3D} \includegraphics[scale=0.19]{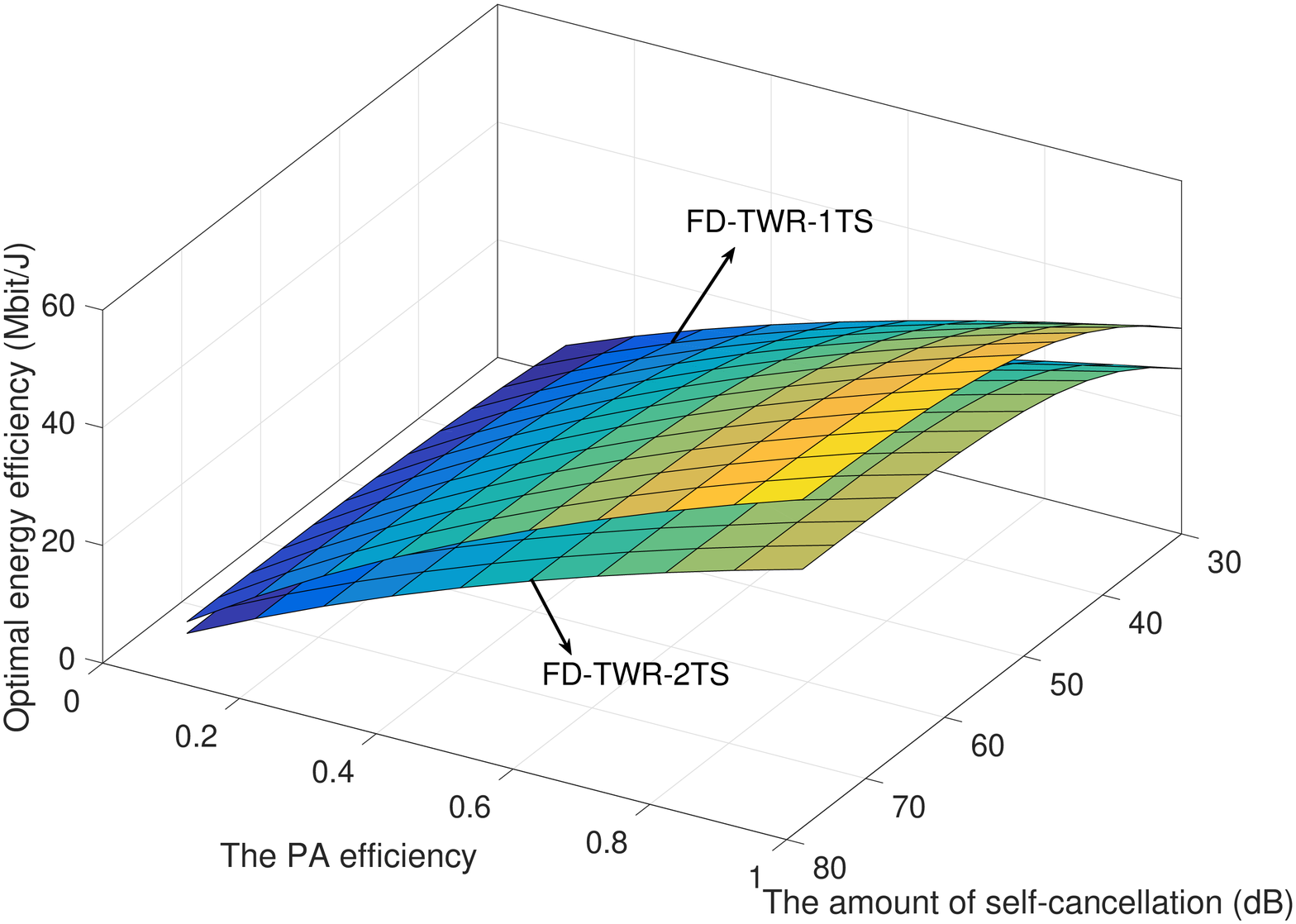}}
\caption{The impact of PA efficiency, self-cancellation and traffic imbalance under ETPA model where $R_{fl} + R_{rl} = 60$ Mbps.}
\label{Ratio_PA_SI_3D}
\end{figure*}
Fig.~\ref{Ratio_PA_SI_3D} shows the impact of PA efficiency, self-cancellation and traffic imbalance on the EE maximization, where the ETPA model is considered for illustration purpose. From Fig.~\ref{Ratio_PA_SI_3D}(a), we can see that FD-TWR can be much more tolerant to the imbalance of traffic than HD-TWR, and the difference of EE between FD-TWR and HD-TWR grows quickly, as the imbalance increases. We also see that FD-TWR-2TS is more robust against the traffic imbalance, compared to FD-TWR-1TS. Particularly, Fig.~\ref{Ratio_PA_SI_3D}(a) shows consistently across the spectrum of PA efficiency that FD-TWR-1TS is suitable for balanced traffic, while FD-TWR-2TS can provide higher EE in the case of severely unbalanced traffic. Fig.~\ref{Ratio_PA_SI_3D}(b) confirms this with consistent results across a wide spectrum of self-cancellation capability. As given earlier, the reason is that FD-TWR-1TS needs to explicitly align the transmission durations of all the nodes. Unbalanced traffic results in excessively high transmit power requirement in some of the links and energy-inefficiently, excessively long transmit duration in the other links, thus compromising the EE.

In addition, Fig.~\ref{Ratio_PA_SI_3D}(c) reveals that the PA efficiency has a dominant effect over the self-cancellation capability, in the case of balanced traffic. This is because a high level of self-cancellation can reduce the the transmit power requirement which is only part of the total PA consumption. In contrast, increasing the PA efficiency can reduce the total PA consumption effectively, thereby resulting in more noticeable EE improvement.
\section{Conclusion}\label{Sec7}
In this paper, the EE maximization of FD-TWR is formulated under practical non-ideal PAs and non-negligible transmission-dependent circuit power. Reformulations of the problems are carried out by deriving the necessary conditions of the optimal solutions, and then rigorously proved to be convex. Extensive simulations show the superiority of FD-TWR in terms of EE when traffic demand is high. The simulations also reveal that the maximum EE of FD-TWR is more sensitive to the PA efficiency than it is to self-cancellation. Also, a full FD design of FD-TWR, {i.e.}, FD-TWR-1TS, is susceptible to unbalanced traffic, while the design with only the relay operating in FD mode, {i.e.}, FD-TWR-2TS, exhibits strong tolerance to traffic imbalance.

\appendices
\section{Proof of The Necessary Condition Of \eqref{EE_Opt_FD_2TS}}\label{Apx1}
\begin{IEEEproof}
We can rewrite the rate constraints in~\eqref{EE_Opt_FD_2TS}, as given by
\begin{equation}\label{rbr-FD2TS}
{C_{a,r}} = \frac{{t_1}}{T}W{\log _2}(1 + \frac{{{P_a}{{\left| {{h_{a,r}}} \right|}^2}}}{{{P_{r,b}}{{\left| {{\tilde{h}_{r,r}}} \right|}^2} + \sigma _r^2}}) \geq {R_{fl}},
\end{equation}
\begin{equation}\label{rru-FD2TS}
{C_{r,b}} = \frac{{t_1}}{T}W{\log _2}(1 + \frac{{{P_{r,b}}{{\left| {{h_{r,b}}} \right|}^2}}}{{\sigma _b^2}}) \geq {R_{fl}},
\end{equation}
\begin{equation}\label{rur-FD2TS}
{C_{b,r}} = \frac{{t_2}}{T}W{\log _2}(1 + \frac{{{P_b}{{\left| {{h_{b,r}}} \right|}^2}}}{{{P_{r,a}}{{\left| {{\tilde{h}_{r,r}}} \right|}^2} + \sigma _r^2}}) \geq {R_{rl}},
\end{equation}
\begin{equation}\label{rrb-FD2TS}
{C_{r,a}} = \frac{{t_2}}{T}W{\log _2}(1 + \frac{{{P_{r,a}}{{\left| {{h_{r,a}}} \right|}^2}}}{{\sigma _a^2}}) \geq {R_{rl}}.
\end{equation}

Assume that~\eqref{EE_Opt_FD_2TS} is feasible, i.e., $P_{\max,i}$, $i\in\{a,b,r\}$, and/or $T$ are sufficiently large to accommodate $R_{fl}$ and $R_{rl}$. We confirm that these constraints are all active at the optimal solution for~\eqref{EE_Opt_FD_2TS}, i.e., equalities are taken in all the constraints. This can be proved by assuming the optimal solution, denoted by $\{P_a^*, P_b^*, P_{r,a}^*, P_{r,b}^*, t_1^*, t_2^*\}$, is inside the feasible solution region where $0\leq P_i^*\leq \Psi^{-1} (P_{\max,i})$ and $0\leq P^*_{r,i}\leq \Psi^{-1} (P_{\max,r})$ for $i\in\{a,b\}$; $0\leq t_1^*+t_2^*\leq T$; and some or all of the constraints \eqref{rbr-FD2TS}--\eqref{rrb-FD2TS} are inactive.

Under this assumption, given $t_1^*$ and $t_2^*$ ($t_1^*+t_2^*\leq T$), we can reduce $P_{r,b}^*$ to ${P_{r,b}^*}'$ and $P_{r,a}^*$ to ${P_{r,a}^*}'$ until equalities can be taken in~\eqref{rru-FD2TS} and~\eqref{rrb-FD2TS}, respectively. We can proceed to reduce $P_a^*$ to ${P_a^*}'$ and $P_b^*$ to ${P_b^*}'$ until equalities are taken in~\eqref{rbr-FD2TS} and~\eqref{rur-FD2TS}, respectively. As a result, the total energy consumption is reduced. Clearly, $0\leq {P_{a}^*}'<P_a^*$, $0 \leq {P_{b}^*}'<P_b^*$, $0\leq {P_{r,a}^*}'<P_{r,a}^*$, and $0 \leq {P_{r,b}^*}'<P_{r,b}^*$; in other words, $\{{P_{a}^*}',{P_{b}^*}',{P_{r,a}^*}',{P_{r,b}^*}',t_1^*,t_2^*\}$ is also a feasible solution for~\eqref{EE_Opt_FD_2TS}. This contradicts with the assumption that $\{P_a^*,P_b^*,P_{r,a}^*,P_{r,b}^*,t_1^*,t_2^*\}$ is the optimal solution. The necessary conditions of the optimal solution for~\eqref{EE_Opt_FD_2TS} are proved.

As a matter of fact, given any pair of feasible $t_1$ and $t_2$ ($t_1+t_2\leq T$), $P_a$, $P_b$, $P_{r,a}$ and $P_{r,b}$ are all minimized (and so is the total energy consumption) if and only if \eqref{rbr-FD2TS}--\eqref{rrb-FD2TS} all take equalities. The minimized powers all satisfy the rate constraints \eqref{rbr-FD2TS}--\eqref{rrb-FD2TS}, but may violate the power constraints of $P_{\max,i}$. In this case, we adjust $t_1$ and $t_2$, until the powers satisfy the power constraints and the total energy consumption is minimized over $\{t_1,t_2\}$, as described in Section~\ref{Sec4}. It is possible that~\eqref{EE_Opt_FD_2TS} is infeasible, that is, for any possible $t_1$ and $t_2$, the minimized powers violate the power constraints; in other words, by no means can the system of interest meet the data rate requirements $R_{fl}$ and/or $R_{rl}$. The data rate requirements need to be reduced, which is beyond the scope of this paper.
\end{IEEEproof}

\section{Proof of The Necessary Condition Of \eqref{EE_Opt_FD_1TS}}\label{Apx2}
\begin{IEEEproof}
The rate constraints in~\eqref{EE_Opt_FD_1TS} can be rewritten as
\begin{equation}\label{rbr-FD1TS}
{C_{a,r}} \hspace{-1mm} = \hspace{-1mm} \frac{t_1}{T}W{\log _2}(\frac{{{P_a}{{\left| {{h_{a,r}}} \right|}^2}}}{{{P_a}{{\left| {{h_{a,r}}} \right|}^2} \hspace{-1mm} + \hspace{-1mm} {P_b}{{\left| {{h_{b,r}}} \right|}^2}}} \hspace{-1mm} + \hspace{-1mm} \frac{{{P_a}{{\left| {{h_{a,r}}} \right|}^2}}}{{{P_r}{{\left| {{\tilde{h}_{r,r}}} \right|}^2} \hspace{-1.5mm} + \hspace{-1mm} \sigma _r^2}}) \hspace{-1mm} \geq \hspace{-1mm} {R_{fl}},
\end{equation}
\begin{equation}\label{rur-FD1TS}
{C_{b,r}} \hspace{-1mm} = \hspace{-1mm} \frac{t_1}{T}W{\log _2}(\frac{{{P_b}{{\left| {{h_{b,r}}} \right|}^2}}}{{{P_a}{{\left| {{h_{a,r}}} \right|}^2} \hspace{-1mm} + \hspace{-1mm} {P_b}{{\left| {{h_{b,r}}} \right|}^2}}} \hspace{-1mm} + \hspace{-1mm} \frac{{{P_b}{{\left| {{h_{b,r}}} \right|}^2}}}{{{P_r}{{\left| {{\tilde{h}_{r,r}}} \right|}^2} \hspace{-1.5mm} + \hspace{-1mm} \sigma _r^2}}) \hspace{-1mm} \geq \hspace{-1mm} {R_{rl}},
\end{equation}
\begin{subequations}\label{rrbu-FD1TS}
\begin{equation}\label{rrb-FD1TS}
{C_{r,a}} = \frac{t_1}{T}W{\log _2}(1 + \frac{{{P_r}{{\left| {{h_{r,a}}} \right|}^2}}}{{{P_a}{{\left| {{\tilde{h}_{a,a}}} \right|}^2} + \sigma _a^2}}) \geq {R_{rl}},\ {\rm or}
\end{equation}
\begin{equation}\label{rru-FD1TS}
{C_{r,b}} = \frac{t_1}{T}W{\log _2}(1 + \frac{{{P_r}{{\left| {{h_{r,b}}} \right|}^2}}}{{{P_b}{{\left| {{\tilde{h}_{b,b}}} \right|}^2} + \sigma _b^2}}) \geq {R_{fl}}.
\end{equation}
\end{subequations}

Assume that~\eqref{EE_Opt_FD_1TS} is feasible, i.e., $P_{\max,i}$, $i\in\{a,b,r\}$, and/or $T$ are sufficiently large to accommodate $R_{fl}$ and $R_{rl}$. We confirm that both \eqref{rbr-FD1TS} and \eqref{rur-FD1TS} are active, and at least one of \eqref{rrb-FD1TS} and \eqref{rru-FD1TS} is active, at the optimal solution for~\eqref{EE_Opt_FD_1TS}. This can be proved by assuming that the optimal solution, denoted by $\{P_a^*, P_b^*, P_{r}^*, t_1^*\}$, is inside the feasible solution region where $0\leq P_i^*\leq \Psi^{-1} (P_{\max,i})$ for $i\in\{a,b,r\}$; $0\leq t_1^*\leq T$; and some or all of the constraints \eqref{rbr-FD1TS}--\eqref{rrbu-FD1TS} are inactive.

Under this assumption, given $t_1^*$ and $P_r^*$, we certainly can reduce $P_a^*$ and $P_b^*$ in an alternating manner until equalities are taken in both~\eqref{rbr-FD1TS} and~\eqref{rur-FD1TS}. Specifically, fixing $P_b^*$ and reducing $P_a^*$ leads to a reduction of $C_{a,r}$ in the left-hand side~(LHS) of~\eqref{rbr-FD1TS} and a growth of $C_{b,r}$ in the LHS of~\eqref{rur-FD1TS}; while fixing $P_a^*$ and reducing $P_b^*$ leads to a reduction of $C_{b,r}$ and a growth of $C_{a,r}$. Nevertheless, this process is surely convergent upon both~\eqref{rbr-FD1TS} and~\eqref{rur-FD1TS} taking the equalities, since $P_a^*$ and $P_b^*$ are not only strictly monotonically decreasing but also lower bounded by
\begin{equation}\label{Apx2_eq2}
\begin{aligned}
P_a^* & \geq \left(2^{\frac{{R_{rl}}T}{W{t_1^*}}} \hspace{-1mm} - \hspace{-1mm} \frac{{{P_a^*}{{\left| {{h_{a,r}}} \right|}^2}}}{{{P_a^*}{{\left| {{h_{a,r}}} \right|}^2} \hspace{-1mm} + \hspace{-1mm} {P_b^*}{{\left| {{h_{b,r}}} \right|}^2}}}\right) \frac{{{P_r^*}{{\left| {{\tilde{h}_{r,r}}} \right|}^2} \hspace{-2mm} + \hspace{-1mm} \sigma _r^2}}{{{\left| {{h_{a,r}}} \right|}^2}} \\
& > \left(2^{\frac{{R_{rl}}T}{W{t_1^*}}} \hspace{-1mm} - \hspace{-1mm} 1 \right) \frac{\sigma _r^2}{{{\left| {{h_{a,r}}} \right|}^2}} > 0,
\end{aligned}
\end{equation}
\begin{equation}\label{Apx2_eq3}
\begin{aligned}
P_b^* & \geq \left(2^{\frac{{R_{fl}}T}{W{t_1^*}}} \hspace{-1mm} - \hspace{-1mm} \frac{{{P_b^*}{{\left| {{h_{b,r}}} \right|}^2}}}{{{P_a^*}{{\left| {{h_{a,r}}} \right|}^2} \hspace{-1mm} + \hspace{-1mm} {P_b^*}{{\left| {{h_{b,r}}} \right|}^2}}}\right) \frac{{{P_r^*}{{\left| {{\tilde{h}_{r,r}}} \right|}^2} \hspace{-2mm} + \hspace{-1mm} \sigma _r^2}}{{{\left| {{h_{b,r}}} \right|}^2}} \\
& > \left(2^{\frac{{R_{fl}}T}{W{t_1^*}}} \hspace{-1mm} - \hspace{-1mm} 1 \right) \frac{\sigma _r^2}{{{\left| {{h_{b,r}}} \right|}^2}} > 0.
\end{aligned}
\end{equation}

Given $P_a^*$ and $P_b^*$ taking equalities in~\eqref{rbr-FD1TS} and~\eqref{rur-FD1TS}, we can proceed to reduce $P_r^*$ until either of~\eqref{rrb-FD1TS} and~\eqref{rru-FD1TS} takes equality. Both ${C_{a,r}}$ and ${C_{b,r}}$ on the LHS of~\eqref{rbr-FD1TS} and~\eqref{rur-FD1TS} grow. Given $P_r^*$, we continue to reduce $P_a^*$ and $P_b^*$ till~\eqref{rbr-FD1TS} and~\eqref{rur-FD1TS} take equalities again, as described above. This process is surely convergent, i.e., \eqref{rbr-FD1TS}, \eqref{rur-FD1TS}, and~\eqref{rrb-FD1TS} or~\eqref{rru-FD1TS}, can all take equalities, since $P_r^*$ is not only strictly monotonically decreasing but also lower bounded by
\begin{equation}\label{Apx2_eq1}
\begin{aligned}
P_r^* \hspace{-1mm} & \geq \hspace{-1mm} \max \hspace{-1mm} \left\{ \hspace{-1mm} (2^{\frac{{R_{rl}}T}{W{t_1^*}}} \hspace{-1mm} - \hspace{-1mm} 1) \frac{{{P_a^*}{{\left| {{\tilde{h}_{a,a}}} \right|}^2} \hspace{-2mm} + \hspace{-1mm} \sigma _a^2}}{{{\left| {{h_{r,a}}} \right|}^2}} ; (2^{\frac{{R_{fl}}T}{W{t_1^*}}} \hspace{-1mm} - \hspace{-1mm} 1) \frac{{{P_b^*}{{\left| {{\tilde{h}_{b,b}}} \right|}^2} \hspace{-2mm} + \hspace{-1mm} \sigma _b^2}}{{{\left| {{h_{r,b}}} \right|}^2}} \hspace{-1mm} \right\} \\
& > \hspace{-1mm} \max \hspace{-1mm} \left\{ \hspace{-1mm} (2^{\frac{{R_{rl}}T}{W{t_1^*}}} \hspace{-1mm} - \hspace{-1mm} 1) \frac{\sigma _a^2}{{{\left| {{h_{r,a}}} \right|}^2}} ; (2^{\frac{{R_{fl}}T}{W{t_1^*}}} \hspace{-1mm} - \hspace{-1mm} 1) \frac{\sigma _b^2}{{{\left| {{h_{r,b}}} \right|}^2}} \hspace{-1mm} \right\}>0.
\end{aligned}
\end{equation}

As a result, all $P_a^*$, $P_b^*$ and $P_r^*$ can be continually reduced to a fixed point where both \eqref{rbr-FD1TS} and \eqref{rur-FD1TS} are active and at least one of \eqref{rrb-FD1TS} and \eqref{rru-FD1TS} is active. Clearly, the fixed point is a feasible solution for~\eqref{EE_Opt_FD_1TS}, and requires less energy than the assumed optimal solution. This contradicts with the assumption that the assumed solution $\{P_a^*,P_b^*,P_r^*,t_1^*\}$ is the optimal feasible solution for~\eqref{EE_Opt_FD_1TS}. The necessary conditions of the optimal solution for~\eqref{EE_Opt_FD_1TS} are proved.

We note that, given any $t_1\leq T$, $P_a$, $P_b$ and $P_{r}$ are all minimized (and so is the total energy consumption) if and only if \eqref{rbr-FD1TS}--\eqref{rrbu-FD1TS} are all active. The minimized powers all satisfy the rate constraints \eqref{rbr-FD1TS}--\eqref{rrbu-FD1TS}, but may violate the power constraints of $P_{\max,i}$. In this case, we adjust $t_1$, until the powers satisfy the power constraints and the total energy consumption is minimized over $t_1 \in [0,T]$, as described in Section~\ref{Sec5}. It is possible that~\eqref{EE_Opt_FD_1TS} is infeasible, that is, for any possible $t_1$, the minimized powers violate the power constraints; in other words, by no means can the system of interest meet the data rate requirements $R_{fl}$ and/or $R_{rl}$. The data rate requirements need to be reduced.
\end{IEEEproof}

\section{Energy Efficiency of HD-TWR-2TS}\label{Apx3}
PNC is also adopted in HD-TWR and two timeslots are required to accomplish TWR. During the first timeslot, nodes $A$ and $B$ transmit signal to node $R$ simultaneously. The received signal at node $R$ can be given by
\begin{equation}\label{yr-HD2TS}
y_r = \sqrt{P_a} \, h_{a,r} \, x_a + \sqrt{P_b} \, h_{b,r} \, x_b + n_r.
\end{equation}

With nested lattice code, the achievable data rates satisfy
\begin{equation}\label{Rbr-HD2TS}
{C_{a,r}} = \frac{{{t_1}}}{T}W{\log _2}(\frac{{{P_a}{{\left| {{h_{a,r}}} \right|}^2}}}{{{P_a}{{\left| {{h_{a,r}}} \right|}^2} + {P_b}{{\left| {{h_{b,r}}} \right|}^2}}} + \frac{{{P_a}{{\left| {{h_{a,r}}} \right|}^2}}}{{\sigma _r^2}}),
\end{equation}
\begin{equation}\label{Rur-HD2TS}
{C_{b,r}} = \frac{t_1}{T}W{\log _2}(\frac{{{P_b}{{\left| {{h_{b,r}}} \right|}^2}}}{{{P_a}{{\left| {{h_{a,r}}} \right|}^2} + {P_b}{{\left| {{h_{b,r}}} \right|}^2}}} + \frac{{{P_b}{{\left| {{h_{b,r}}} \right|}^2}}}{{\sigma _r^2}}).
\end{equation}

During the second timeslot, node $R$ forwards the combined signal $x_r$ to nodes $A$ and $B$. Thus, the received signal at nodes $A$ and $B$ can be given by
\begin{equation}\label{yi-HD2TS}
{y_i} = \sqrt {{P_r}} {h_{r,i}}{x_r} + {n_i},\ i \in \left\{ {a,b} \right\}.
\end{equation}

By cancelling the self-interfere, the achievable data rates at nodes $A$ and $B$ in the second timeslot satisfy~\cite{latticecode}
\begin{equation}\label{Rrb-HD2TS}
{C_{r,a}} = \frac{{{t_2}}}{T}W{\log _2}(1 + \frac{{{P_r}{{\left| {{h_{r,a}}} \right|}^2}}}{{\sigma _a^2}}),
\end{equation}
\begin{equation}\label{Rru-HD2TS}
{C_{r,b}} = \frac{{{t_2}}}{T}W{\log _2}(1 + \frac{{{P_r}{{\left| {{h_{r,b}}} \right|}^2}}}{{\sigma _b^2}}).
\end{equation}

The EE maximization problem has the same form as that of FD-TWR-2TS, but the total energy consumption is different, as given by
\begin{align}\label{ETotal_HD_2TS}
{E_{{\rm total}}} &= \left( {{P_{tx,a}} + {P_{tx,b}} + {P_{rx,r}}} \right){t_1} \\
&+ \left( {{P_{rx,a}} + {P_{rx,b}} + {P_{tx,r}}} \right){t_2} + {P_{idle}}\left( {T - {t_1} - {t_2}} \right), \nonumber
\end{align}
where ${P_{idle}} = {P_{idle,a}} + {P_{idle,b}} + {P_{idle,r}}$. The necessary conditions can also be given by \eqref{Equal_RCon1} and \eqref{Equal_RCon2}, as can be proved in the same way as FD-TWR-1TS. The optimal transmit powers, $P_a$, $P_b$ and $P_r$, are written as
\begin{equation}\label{Pb_HD_2TS}
{P_a}\left( {{t_1}} \right) = \left( {{\lambda _1} - \frac{{{\lambda _1}}}{{{\lambda _1} + {\lambda _2}}}} \right)\frac{{\sigma _r^2}}{{{{\left| {{h_{a,r}}} \right|}^2}}},
\end{equation}
\begin{equation}\label{Pu_HD_2TS}
{P_b}\left( {{t_1}} \right) = \left( {{\lambda _2} - \frac{{{\lambda _2}}}{{{\lambda _1} + {\lambda _2}}}} \right)\frac{{\sigma _r^2}}{{{{\left| {{h_{b,r}}} \right|}^2}}},
\end{equation}
\begin{equation}\label{Pr_HD_2TS}
{P_r}\left( {{t_2}} \right) = \max \left\{ {\left( {{\lambda _3} - 1} \right)\frac{{\sigma _b^2}}{{{{\left| {{h_{r,b}}} \right|}^2}}},\left( {{\lambda _4} - 1} \right)\frac{{\sigma _a^2}}{{{{\left| {{h_{r,a}}} \right|}^2}}}} \right\},
\end{equation}
where ${\lambda _1} = {2^{\frac{{{R_{fl}}T}}{{W{t_1}}}}}$, ${\lambda _2} = {2^{\frac{{{R_{rl}}T}}{{W{t_1}}}}}$, ${\lambda _3} = {2^{\frac{{{R_{fl}}T}}{{W{t_2}}}}}$ and ${\lambda _4} = {2^{\frac{{{R_{rl}}T}}{{W{t_2}}}}}$ for notational simplicity.

By substituting~\eqref{Pb_HD_2TS}--\eqref{Pr_HD_2TS} into the optimization of HD-TWR-2TS, we can obtain the equivalent objective function under the TPA model, as given by
\begin{align}
E(t_1,t_2) &= [\sqrt{\frac{P_{\max ,a}}{\eta _{\max ,a}^2}} \sqrt{P_a(t_1)} \hspace{-1mm} + \hspace{-1mm} \sqrt{\frac{P_{\max ,b}}{\eta _{\max ,b}^2}} \sqrt{P_b(t_1)} \hspace{-1mm} + \hspace{-1mm} P_1]t_1 \nonumber \\
&+ [\sqrt{\frac{P_{\max ,r}}{\eta _{\max ,r}^2}} \sqrt{P_r(t_2)} + P_2]t_2  \nonumber \\
&+({P_{idle,a}} + {P_{idle,b}} + {P_{idle,r}})T, \nonumber
\end{align}
where ${P_1} = {P_{rx,r}} + P_{base,a} + {P_{base,b}} + \varepsilon (R_{fl}+R_{rl}) - {P_{idle}},\ {P_2} = {P_{rx,a}} + {P_{rx,b}} + {P_{base,r}} + \varepsilon \max\{R_{fl},R_{rl}\} - {P_{idle}}$; and the equivalent objective function under the ETPA model, as formulated by
\begin{align}
E(t_1,t_2) &= [\frac{P_a(t_1)}{(1 + u \kappa_a)\eta _{\max ,a}} + \frac{P_b(t_1)}{(1 + u \kappa_b)\eta _{\max ,b}} + P_1]t_1 \nonumber \\
&+ [\frac{P_r(t_2)}{(1 + u \kappa_r)\eta _{\max ,r}} + P_2]t_2 + {P_{idle}}T, \nonumber
\end{align}
where ${P_1} = \frac{{u \kappa_a {P_{\max ,a}}}}{{\left( {1 + u \kappa_a} \right){\eta _{\max ,a}}}} + \frac{{u \kappa_b {P_{\max ,b}}}}{{\left( {1 + u \kappa_b} \right){\eta _{\max ,b}}}} + {P_{rx,r}} + P_{base,a} + {P_{base,b}} + \varepsilon (R_{fl}+R_{rl}) - {P_{idle}}$ and ${P_2} = \frac{{u \kappa_r {P_{\max ,r}}}}{{\left( {1 + u \kappa_r} \right){\eta _{\max ,r}}}} + {P_{rx,a}}  + {P_{rx,b}} + {P_{base,r}} + \varepsilon \max\{R_{fl},R_{rl}\} - {P_{idle}}$.

It can be proved that $E(t_1,t_2)$ is quasi-convex under the TPA model and convex under the ETPA model. The detailed proofs are suppressed due to the limited space.

\ifCLASSOPTIONcaptionsoff
  \newpage
\fi

\end{document}